\documentclass{ws-rv975x65}  
\usepackage{epsfig}  
\usepackage{bm}
\newcommand{\Fd}{F^{\dagger}}

\newcommand{\ep}{\varepsilon}
\def\Re{\mathop{\rm Re}}

\newcommand {\vecq}{\bm q}
\newcommand {\vecr}{\bm r}
\newcommand {\vecP}{\bm P}

\newcommand {\vecp}{\bm p}

\newcommand {\vecn}{\bm n}
\newcommand {\vecv}{\bm v}
\newcommand {\vecnabla}{\bm \nabla}

\newcommand {\vecsigma}{\bm \sigma}
\newcommand {\vectau}{\bm \tau}
\newcommand {\vecnu}{\bm \nu}
\newcommand{\vecphi}{\bm \phi}
\newcommand{\bea}{\begin{eqnarray}}
\newcommand{\eea}{\end{eqnarray}}
\newcommand{\be}{\begin{equation}}
\newcommand{\ee}{\end{equation}}
\begin{document}
\setcounter{chapter}{0}

\chapter{Nuclear Superconductivity in Compact Stars: \\
BCS Theory and Beyond}

\author{Armen~Sedrakian}
\address{Institute for Theoretical Physics, T\"ubingen 
University, 72076 T\"ubingen, Germany
}

\author{John W.~Clark}
\address{Department of Physics, Washington University, St. Louis,
Missouri 63130, USA}

\markboth{A. Sedrakian and J. W. Clark}
{Nuclear superconductivity in compact stars}


\begin{abstract}
This chapter provides a review of microscopic theories of pairing in 
nuclear systems and neutron stars.  Special attention is
given to the mean-field BCS theory and its extensions to 
include effects of polarization of the medium and retardation 
of the interactions.  Superfluidity in nuclear systems that
exhibit isospin asymmetry is studied.  We further address
the crossover from the weak-coupling BCS description to the 
strong-coupling BEC limit in dilute nuclear systems.  Finally, 
within the observational context of rotational anomalies of 
pulsars, we discuss models of the vortex state in superfluid 
neutron stars and of the mutual friction between superfluid and 
normal components, along with the possibility of type-I 
superconductivity of the proton subsystem.
\end{abstract}

\section{Introduction}     

Neutron stars represent one of the densest concentrations of matter
in our universe.  These compact stellar objects are
born in the gravitational collapse of luminous stars 
with masses exceeding the Chandrasekhar mass limit.  The 
observational phenomena characteristic of neutron stars,
such as the pulsed radio emission, thermal X-ray 
radiation from their surfaces, and gravity waves emitted 
in isolation or from binaries, provide information on 
their structure, composition, and dynamics.
The properties of superdense matter are fundamental to 
our understanding of nature at small distances characteristic 
of nuclear forces and of the underlying theory 
of strong interactions -- QCD.   In fact, neutron stars (NS)
provide a unique setting in which all of the known forces 
-- strong, electroweak, and gravitational -- play essential 
roles in determining observable properties.\index{neutron stars}
\index{pulsars}

Except for the very early stages of their evolution, neutron stars 
are extremely cold, highly degenerate objects.  Their interior 
temperatures are typically a few hundreds of keV, far below the 
characteristic Fermi energies of the constituent fermions, which 
run to tens or hundreds of MeV.  
Although very repulsive at short distances, the strong
interaction between nucleons is sufficiently attractive to
induce a pairing phase transition to a superfluid state
of neutron-star matter.  As will be discussed in this chapter, 
the existence of superfluid components within neutron stars
has far-reaching consequences for their observational manifestations. 

Historically, the first observational evidence for superfluidity 
in neutron-star interiors was provided by the timing of radio 
emissions of pulsars, the first class of neutron stars,
discovered in 1967 by Jocelyn Bell.  The pulsed emissions, with
a typical periodicity of seconds or less, are locked to the 
rotation period of the star.  Although pulsars are nearly prefect 
clocks, their periods increase gradually over time, corresponding
to a secular loss of rotational energy.  Significantly, some 
pulsars are found to exhibit deviations from this impressive
regularity.  The pulsar timing anomalies divide roughly into
three types. (i)~{\sl Glitches or Macrojumps}.  These are
distinguished by abrupt increases in the rotation and spin-down 
rates of pulsars by amounts $\Delta\Omega/\Omega \sim 10^{-6}-10^{-8}$ and 
$\Delta\dot \Omega/\Omega \sim 10^{-3}$.  After a glitch, 
$\Delta\Omega/\Omega$ and $\Delta\dot \Omega/\Omega$ slowly
relax toward their pre-glitch values, on a time scale of order
weeks to years, in some cases with permanent hysteresis 
effects.\cite{GLITCHES,TIMING_NOISE}
\index{pulsars!glitches}
Such behavior is attributed to a component within the star 
that is only weakly coupled to the rigidly rotating normal component
responsible for the emission of pulsed radiation -- an interpretation 
supported by fits of the measured rotation 
under different modeling assumptions.  (ii)~{\sl Timing Noise or Microjumps}. 
These represent irregular,  stochastic deviations in 
the spin and spin-down rates that are superimposed on the 
near-perfect periodic rotation of the star.  The origin of 
microjumps remains unclear, but they could be evidence of 
stochastic coupling between the superfluid and normal 
components.\cite{TIMING_NOISE} 
(iii)~{\sl Long-Term Periodic Variabilities}.  
Observed in the timing of few pulsars, most notably PSR 
B1828-11, these deviations strongly constrain theories of 
superfluid friction inside NS, if their periodicities are 
interpreted in terms of NS precession.\cite{PRECESSION1} 
\index{pulsars!precession}

X-ray observations from orbiting spacecraft, which yield 
estimates of surface temperature for a half-dozen or so young 
neutron stars, further reinforce the picture of NS with superfluid 
content.\cite{XRAYS1,XRAYS2,XRAYS3,XRAYS4}  
At the stellar ages involved, neutrino emission from 
the dense interior dominates thermal evolution, with nucleonic 
superfluidity acting to suppress the main emission mechanisms.  
The existing measurements of surface temperatures indicate 
that superfluid hadronic components must be present in some 
NS, since otherwise they would cool to temperatures below 
the empirical estimates on very short time scales.  Finally, 
since NS with their huge gravitational fields are expected to
be major sources of gravitational wave radiation, it is believed 
that observation of gravitational waves from oscillating neutron 
stars can provide further information on the state of matter in
their interiors.  In particular, the eigenfrequencies and damping 
rates of gravity waves may carry imprints of dissipation processes 
in the superfluid phases.\cite{GRAVITY1,GRAVITY2,GRAVITY3}

\begin{figure}[t]
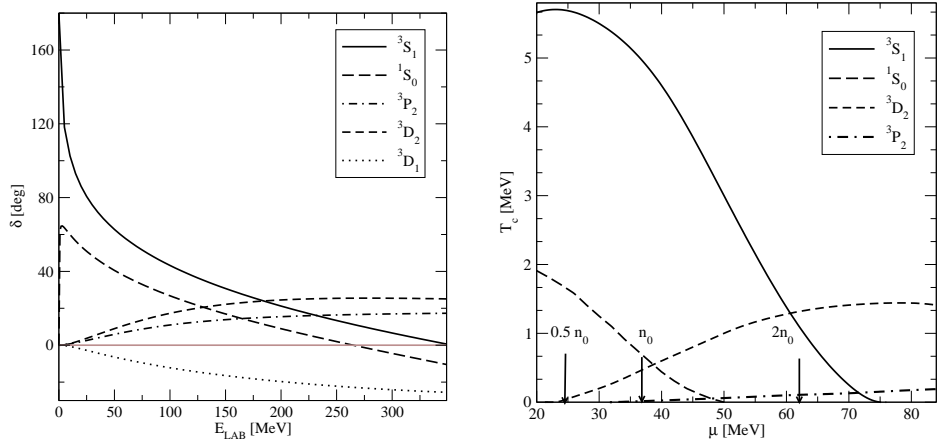

\centerline{
\psfig{file=fp_fig1a.eps,width=2.3in,angle=0}\hspace{.55cm}
\psfig{file=fp_fig1b.eps,width=2.3in,angle=0}
}
\vskip 0.2cm
\caption{{\it Left panel.} Dependence of experimental scattering 
phase shifts in $^3S_1$, $^3P_2$, $^3D_2$, and $^3D_1$ partial waves 
on laboratory energy. {\it Right panel.} Dependence of 
critical temperatures of superfluid phase transitions in attractive 
channels on chemical potential.  The corresponding densities are 
indicated by arrows.  \label{fig:TC}}
\end{figure}

The existence of neutron-star superfluidity, first envisioned
by Migdal~\cite{MIGDAL} in 1959, is broadly consistent with 
microscopic theories of nucleonic matter in NS.  Shortly after the 
advent of the Bardeen-Cooper-Schrieffer (BCS) theory in 1957, 
BCS pairing of nucleons in nuclei and infinite nuclear matter was 
suggested and
studied.\cite{PINES,COOPER} With the discovery of pulsars,
the implications of nucleonic pairing for neutron-star properties 
were explored,\cite{BAYM_NATURE} and viable microscopic calculations
of pairing gaps began to appear soon 
thereafter.\cite{YANG1,YANG2,CHAO,HOFFBERG70,TAKATSUKA72}

Partial-wave analysis of the nucleon-nucleon ($NN$) scattering data 
yields information on the dominant pairing channel in nuclear- and 
neutron-matter problems in a given range of density (see 
Fig.~\ref{fig:TC}).  At high densities, corresponding to laboratory 
energies above 250 MeV, the most attractive pairing channel is 
the tensor-coupled $^3P_2$--$^3F_2$ channel,\cite{P_WAVE0,P_WAVE} 
whenever isospin symmetry is even slightly broken. 
This condition holds inside neutron stars, with the partial densities
of neutron and proton fluids differing quite significantly, except 
in special meson-condensed phases where the nucleonic matter is 
isospin-symmetric.  \index{phase shifts} In such a case the 
phase-shift analysis predicts that the most attractive pairing 
interaction is in the $^3D_2$ wave.\cite{D_WAVE1,D_WAVE2} 

At low density, isospin-symmetric nuclear matter exhibits pairing 
due to the attractive interaction in the $^3S_1$--$^3D_1$ partial wave, 
a tensor component of the force again being responsible for 
the coupling of the $S$ and $D$ 
waves.\cite{SDPAIRING1,SDPAIRING2,SDPAIRING3,SDPAIRING4}  This 
interaction channel is distinguished by its ability to 
support a two-body bound state in free space -- the deuteron.\index{deutron} 
However, under the highly isospin-asymmetric condition typical in
neutron stars, neutron-proton pair condensation is quenched by the
large discrepancy between the neutron and proton Fermi momenta 
(see Sec.~\ref{SECTION3}). 

This review is devoted to several aspects of nucleonic superfluidity 
in neutron stars that are of major current interest. 
Section~\ref{SECTION2} provides an overview of the many-body theories
of pairing in neutron stars, with a special focus on the Green's 
function description of pairing, and the effects of self-energies 
and vertex corrections.  The key ideas of the correlated-basis
(or ``CBF'') approach to superfluid states of Fermi systems
are also presented.
Section~\ref{SECTION3} is concerned with the possibility of
pairing between particles lying on different Fermi surfaces,
in particular, between protons and neutrons in asymmetric nuclear 
matter.  Here we determine the critical value of isospin asymmetry at 
which a transition from $^3S_1$--$^3D_1$ to $^1S_0$ pairing occurs. 
(This threshold value is small compared to $p-n$ asymmetries 
typical of neutron-star matter).  More generally, we consider
several competing phases that could exist in asymmetric mixtures
of fermion species.  There exist systems of this kind, notably 
dilute, ultracold atomic gases and baryonic matter as described
by QCD at high density, where ``asymmetric pairing'' is enforced.
(In the case of dilute atomic gases, such conditions can be
``tuned in'' by external fields, while in the QCD problem, 
the conditions of charge neutrality and $\beta$ equilibrium
together with the heaviness of the strange quark lead naturally
to asymmetric pairing among light quarks.)  In Sec.~\ref{SECTION4} 
we turn to the phenomenon of crossover from BCS superconductivity 
to Bose-Einstein condensation, as it occurs in fermionic systems that 
support a two-body bound state in free space.  Section~\ref{SECTION5} is 
devoted to the physics of superfluids at the ``mesoscopic'' scale, 
with discussions of flux quantization, neutron vorticity, the
electrodynamics of superconducting protons, and their implications 
for modeling the dynamics of rotational anomalies in pulsar 
timing. Our conclusions and related perspectives are summarized in 
Sec.~\ref{SECTION6}. 
Two other recent reviews\cite{R1,R2} offer complementary 
information and perspectives on nuclear pairing and nucleonic 
superfluidity.

\section{ Many-body theories of pairing}
\label{SECTION2}

\subsection{Propagators}
\label{GFA}

This subsection outlines the Green's functions method for the treatment 
of superfluid systems.  The original formulations of this approach are
due to Gor'kov and Nambu, who employed thermodynamic Green's 
functions.\cite{ABRIKOSOV}  
Herein we consider the real-time, finite-temperature formalism, 
which is suited to studies of both equilibrium and non-equilibrium 
systems.  Our discussion is restricted to equilibrium systems, and we shall 
work with the retarded components of the full Green's function of the
non-equilibrium theory. Superfluid systems are described in 
terms of  $2\times 2$ matrices of propagators (known as Nambu-Gor'kov 
matrix propagators) that are defined as \index{Nambu-Gorkov propagators}
\begin{eqnarray} 
\label{GF1}
{\cal G}_{\alpha\beta}(x,x') &=& 
\left( \begin{array}{cc}
G_{\alpha\beta}(x,x') & F_{\alpha\beta}(x,x')\\
-F^{\dagger}_{\alpha\beta}(x,x') & \tilde G_{\alpha\beta}(x,x')\\
\end{array}
\right)\nonumber\\& =& 
\left( \begin{array}{cc}
-i\langle T\psi_{\alpha}(x)\psi_{\beta}^{\dagger}(x')\rangle 
&\langle T\psi_{\alpha}(x)\psi_{\beta}(x')\rangle \\
\langle T\psi_{\alpha}^{\dagger}(x)\psi_{\beta}^{\dagger}(x')\rangle
&-i\langle T\psi^{\dagger}_{\alpha}(x)\psi_{\beta}(x')\rangle
\end{array}
\right)\,,
\end{eqnarray}
where $\psi_{\alpha}(x)$ are the baryon field operators, $x$ is the
space-time coordinate,  the indices $\alpha$ and $\beta$ stand for 
the internal (discrete)
degrees of freedom, and $T$ and $\tilde T$ denote time-ordering and inverse 
time-ordering of operators, respectively. 
The $2\times 2$ matrix Green's function (\ref{GF1}) satisfies the 
Schwinger-Dyson equation \index{Dyson-Schwinger equations} 
\begin{eqnarray}\label{DYSON}
{\cal G}_{\alpha\beta}(x,x') = {\cal G}^0_{\alpha\beta}(x,x') 
+ \sum_{\gamma , \delta}\!\int\!\!d^4x'' d^4x'''
{\cal G}^0_{\alpha\gamma}(x,x'''){\Omega}_{\gamma\delta}(x''',x'') 
{\cal G}_{\delta\beta} (x'',x'),
\end{eqnarray}
where the free-propagator matrices ${\cal G}^0_{\alpha\beta}(x,x')$
are diagonal in the Nambu-Gor'kov space.
The matrix structure of the self-energy $\Omega_{\alpha\beta}(x,x')$ 
is identical to that of the propagators: the on-diagonal elements are 
$\Sigma(p)$ and $\Sigma(-p)$, and the off-diagonal elements are 
$\Delta(p)$ and $\Delta^{\dagger}(p)$.  Fourier transforming Eq.~(\ref{DYSON}) 
with respect to the relative coordinate $x-x'$, one obtains the Dyson
equation in the momentum representation.  In this representation, 
the components of the Nambu-Gor'kov matrix obey the 
coupled Dyson equations 
\begin{eqnarray}\label{1}
{ G}_{\alpha\beta}(p) &=& { G}_{0\alpha\beta}(p) + 
{ G}_{0\alpha\gamma}(p) \left[{ \Sigma}_{\gamma\delta}(p) 
{ G}_{\delta\beta}(p)
+{ \Delta}_{\gamma\delta}(p){\Fd}_{\delta\beta}(p) \right]\,,\\
\label{2}
{ \Fd}_{\alpha\beta} (p) &=& { G}_{0\alpha\gamma}(-p)\left[
{ \Delta}^{\dagger}_{\gamma\delta}(p) { G}_{\delta\beta}(p) 
+{ \Sigma}_{\gamma\delta}(-p){ \Fd}_{\delta\beta} (p) \right]\,.
\end{eqnarray}
Here ${ G}_{\alpha\beta}(p)$ and ${ G}_{0\alpha\beta}(p)$ are 
the full and free normal propagators, ${ \Fd}_{\alpha\beta} (p)$ and  
${ F}_{\alpha\beta} (p)$ are the anomalous propagators, and 
${ \Sigma}_{\alpha\beta}(p)$ and ${ \Delta}_{\alpha\beta}(p)$  
are the normal and anomalous self-energies.  The Greek subscripts 
are the spin/isospin indices, and summation over repeated indices 
is understood. For systems with time-reversal symmetry, it is sufficient 
to solve Eqs.~(\ref{1}) and (\ref{2}), since this symmetry implies 
that 
$
\Delta_{\alpha\beta}(p)=[ \Delta^{\dagger}_{\alpha\beta}(p)]^* .
$
It is instructive to rewrite Eqs.~(\ref{1})--(\ref{2}) 
in terms of  auxiliary Green's functions
\begin{eqnarray} \label{N1}
G^N_{\alpha\beta}(p) &=&  G_{0\alpha\beta}(p)+G^N_{\alpha\gamma}(p)
\Sigma_{\gamma\delta}(p) G_{0\delta\beta}(p)
\end{eqnarray}
describing the unpaired state.
The solution of this equation is $G^N_{\alpha\beta}(p) 
= \delta_{\alpha\beta}[\omega -\varepsilon(p)]^{-1}$, where 
$\varepsilon(p) = \epsilon_p + \Sigma(p)$ and $\epsilon_p$ 
is the free single-particle spectrum. (N.B.\ Assuming 
that the forces conserve spin and isospin, the self-energy 
$\Sigma(p)$ is diagonal in spin and isospin spaces).
Combining Eqs.~(\ref{1}), (\ref{2}), and (\ref{N1}), we
derive an alternative but equivalent form of the Schwinger-Dyson 
equations, namely 
\begin{eqnarray}\label{D1}
 G_{\alpha\beta}(p) =  G^N_{\alpha\gamma}(p)
\left[\delta_{\gamma\beta}+\Delta_{\gamma\delta}(p)
\Fd_{\delta\beta}(p)\right],\quad 
\label{D2}
\Fd_{\alpha\beta}(p) =  G^N_{\alpha\gamma}(-p) 
\Delta^{\dagger}_{\gamma\delta}(p)  G_{\delta\beta}(p)\, ,\nonumber 
\end{eqnarray}
which can be solved to obtain
\bea \label{GF2}
G_{\alpha\beta}(p) &=& \delta_{\alpha\beta} 
\frac{\omega-E_A(p)+E_S(p)}
{\left[\omega-E_A(p)\right]^2-E_S(p)^2-\Delta^2(p)}\,,\\  
\label{GF3}
\Fd_{\alpha\beta}(p) &=& 
\frac{\Delta^{\dagger}_{\alpha\beta}(p)}
{\left[\omega-E_A(p)\right]^2-E_S(p)^2-\Delta^2(p)}\,.
\eea
Here we have made the substitution 
$\Delta(p)\Delta^{\dagger}(p) = -\Delta^2(p)$
and defined symmetric and antisymmetric parts of the single-particle
spectrum in the normal state, 
$E_{S/A} = \left[\varepsilon(p)\pm\varepsilon(-p)\right]/2.$
For isotropic systems, the self-energy is invariant under reflections
in space (i.e.\ the self-energy is even under $\vecp \to -\vecp)$;
furthermore, for systems that are time-reversal invariant, the 
self-energies are even under the transformation $\omega \to -\omega$.
Hence the antisymmetric piece of the spectrum $E_{A}$ must
be absent when both conditions are met.
The poles of the propagators (\ref{GF2}) determine the excitation
spectrum of the superfluid system, given by
\be \label{BRANCHES}
\omega_{\pm} = E_A(p)\pm \sqrt{E_S(p)^2 + \Delta^2(p) }\,.
\ee
Here one sees that there is a finite energy cost $\sim 2\Delta$ for 
creating an excitation from the ground state of the system when 
$E_A = 0$, a property that leads to the existence of superflow 
or supercurrent in paired fermionic systems. 

The solutions (\ref{GF2}) and (\ref{GF3}) are completely general, all 
functions being dependent on the three-momentum and the energy.
Superfluid systems are often treated in the quasiparticle approximation, 
\index{quasiparticle approximation}
in which the self-energies are approximated by their on-mass-shell 
counterparts. The rationale behind such an approach is that the nuclear 
system in its ground state can then be described in terms of 
Fermi-liquid theory, if one neglects pair correlations.  Switching 
on the pair correlations precipitates a rearrangement of the 
Fermi-surface, but it is assumed that the quasiparticle 
concept remains intact.  \index{Fermi liquid}

Within this framework, the wave-function renormalization 
\index{wave-function renormalization}
is defined by expanding the normal self-energy as 
$
\Sigma(\omega) = \Sigma(\varepsilon_p) + \partial_{\omega}
\Sigma(\omega)\vert_{\omega = \varepsilon_p} (\omega-\varepsilon_p)
$,  
where $\varepsilon_p = \epsilon_p + {\rm Re}\,\Sigma (\varepsilon_p)$ 
is the on-mass-shell single-particle spectrum in the normal state 
that solves Eq.~(\ref{N1}). Now, it is seen that the propagators
(\ref{GF2}) and (\ref{GF3}) retain their form if they are 
renormalized as  
\be\label{RENORMALIZATION}
\tilde G(p) =  Z(\vecp)G(\omega+i\delta,\vecp), 
\quad \tilde \Fd(p) =  Z(\vecp)\Fd(\omega+i\delta,\vecp), \quad 
\tilde \Delta^2(p) =  Z(\vecp)^{2} \Delta^2(\vecp)\,,
\ee
where 
$Z(\vecp) \equiv \left[1-\partial_{\omega} 
\Sigma(\omega)\vert_{\omega = \varepsilon_p}\right]^{-1}$ 
is the wave-function renormalization  
and the tilde identifies renormalized quantities.
An additional feature of the renormalized propagators is that the 
quasiparticle spectrum 
$\varepsilon(p)$ is now constrained to the mass shell.
(N.B.\ For time-local interactions the gap function is energy-independent, 
so there is no need to expand $\Delta(\omega)$ around its on-shell value. 
We shall return to the off-shellness of the self-energies  
in Subsec.~\ref{ELIASHBERG}.)

Renormalization of propagators within the quasiparticle picture suggests 
that the probability of finding an excitation with given momentum $\vecp$ 
is strongly peaked at the value $\varepsilon_p$.  The wave-function 
renormalization takes into account corrections that are linear in the 
departure from this value.  From the computational point of view, such 
corrections require a knowledge of the off-shell normal self-energies. 
The dependence of the imaginary part of the self-energy (quasiparticle 
damping) on frequency follows from Fermi-liquid theory, being 
given by $2\,{\rm Im}\Sigma (\omega) = a \left[ 1 + \left({\omega}/
{2\pi T}\right)^2\right]$, where $a$ is a density-dependent constant. 
The real part of the self-energy, ${\rm Re}\Sigma(\omega)$, can be 
computed from ${\rm Im}\Sigma (\omega)$ via the Kramers-Kronig dispersion 
relation only if the latter function is known for all frequencies $\omega \in 
[-\infty, \infty]$.  Accordingly, the foregoing result from Fermi-liquid 
theory should be supplemented by a model of the high-energy tail of the 
quasiparticle damping.

The momentum dependence of propagators can be approximated by introducing 
an effective quasiparticle mass. For nonrelativistic particles, 
expansion of the normal self-energy around the Fermi momentum leads to
\be
\varepsilon(p) = \frac{p_F}{m^*}(p-p_F) - \mu^*\,,\quad \quad
\frac{m^*}{m}  = \left[1+\frac{m}{p_F}
\partial_{p}\Re\Sigma(p)\vert_{p=p_F}\right]^{-1}\,.
\ee
where $\mu^*\equiv -\epsilon(p_F)+\mu - \Re \Sigma(p_F)$. 
A closed system of equations determining the properties of the
superfluid system is obtained by specifying the self-energies 
in terms of the propagators and interactions, as discussed
below.

\subsection{Mean-field BCS theory}
\label{BCS}

The BCS-type theory of superconductivity as applied to nuclear systems
is predicated on a mean-field approximation to the anomalous
self-energy.  Specifically, the anomalous self-energy is expressed
through the four-point vertex function $\Gamma(p,p')$
in the form
\be
\Delta(p) = -2\int\!\frac{d^4p'}{(2\pi)^4} \Gamma(p,p')
\,{\rm Im}F(p')f(\omega')\,,
\label{ASEMF}
\ee
where $f(\omega) = [1+{\rm exp}(\beta\omega)]^{-1}$ and $\beta$ is
the inverse temperature.   We observe that the gap is energy-independent 
when the interactions are local in time, corresponding to no retardation, 
as in the case where the effective interaction 
$\Gamma (p,p')$ is replaced by the bare interaction $V(\vecp,\vecp')$.
In this case, carrying out the renormalization according to 
Eq.~(\ref{RENORMALIZATION}) and integrating over the energy, 
we arrive at\cite{MIGDAL_TFFS,BALDO_Z,LOMBARDO_Z}
\be \label{GAP_ONSHELL}
\tilde \Delta(\vecp) = Z(\vecp)\int\!\frac{d^3 p'}{2(2\pi)^3} 
V(\vecp,\vecp')Z(\vecp')\frac{\tilde \Delta(\vecp')}
{\tilde\omega_+(\vecp')}
\left[f(\omega_+)-f(\omega_-)\right]\,,
\ee
where $\tilde\omega_{\pm} = \pm\sqrt{\varepsilon_p^2 + \tilde\Delta^2}$ 
[cf.\ Eq.~(\ref{BRANCHES})]. 
Further progress requires partial-wave decomposition of the 
interaction in Eq.~(\ref{GAP_ONSHELL}). To avoid excessive notation,
our consideration focuses on a single, uncoupled channel, for which
we obtain a one-dimensional gap equation
\be\label{GAP_PARTIAL}
\tilde\Delta(p)=  Z(p)\int \frac{dp \,p^2}{(2\pi)^2} V(p,p')Z(p')
\frac{\tilde\Delta(p')}{\tilde\omega_+(p')}
\left[f(\tilde\omega_+)-f(\tilde\omega_-)\right]\,,
\ee
where $V(p,p')$ is the interaction in the given partial wave.
The gap equation is supplemented by the equation for the density 
of the system, 
\bea \label{DENS1}
\rho = -2g\int\!\! \frac{d^4 p}{(2\pi)^4} {\rm Im}G(p) f(\omega)
= \frac{g}{2}\int\!\! \frac{d^3 p}{(2\pi)^3}Z(\vecp)
       \sum_{i = +, -}
\left(1 + \frac{\varepsilon_p}{\tilde\omega_i}\right)f(\tilde\omega_i)\,,
\eea
which determines the chemical potential in a self-consistent manner.
Here $g$ is the isospin degeneracy factor.   
Eqs.~(\ref{ASEMF})--(\ref{DENS1}) define mean-field BCS theory.  
In subsequent discussions based on this theory, the tilde notation 
will be suppressed.

Assuming that the interaction $V(p,p')$ and the single-particle 
spectrum $\varepsilon_p$ in the unpaired state are known, 
Eqs.~(\ref{GAP_PARTIAL}) and (\ref{DENS1}) form a closed system 
for the gap and chemical potential.
Within this formulation, the presence of a hard core in the
interaction causes no overt problems, for one can readily show
that at the critical temperature, Eq.~(\ref{GAP_PARTIAL}) transforms 
into an integral equation that sums the particle-particle ladder
series to all orders.
In the special case in which the interaction is momentum-independent,
Eq.~(\ref{GAP_PARTIAL}) leads to the familiar weak-coupling formula
\be \label{WCF}
\Delta(p_F) \simeq 8\mu^* \, {\rm exp}
\left(-\frac{1}{\nu(p_F)|V(p_F,p_F)|}\right)\,,
\ee
where the density of states is specified by
$\nu(p_F) = m^*p_F Z^2(p_F)/2\pi^2$ (for one direction of isospin). 
The weak-coupling formula is often used to estimate the magnitude
of the gap.  However, because realistic pairing interactions are
momentum-dependent, the density dependence of the 
gap predicted by Eq.~(\ref{GAP_PARTIAL}) may deviate markedly 
from that given by the weak-coupling formula (\ref{WCF}).
Moreover, as argued in Ref.~\refcite{KKC}, if the pairing interaction
acquires a strong momentum dependence due to a short-range
repulsive core, the weak-coupling formula \index{BCS weak-coupling formula}
may well produce a meaningless or useless estimate of the gap, since its
derivation requires that $V(p,p')$ take a negative value
on the Fermi surface.

The normal-state self-energy is written as 
\be\label{SIGMA}
\Sigma (p) = -2\int\frac{d^4p'}{(2\pi)^4} T
\left(p,p'; p+p'\right)_A {\rm Im}G(p')f(\omega')\,,
\ee
where the subscript $A$ indicates antisymmetrization of the final 
states and the contribution $\propto {\rm Im}T$ is omitted for 
simplicity. For nuclear systems, the amplitude $T$ is often 
approximated by the scattering $T$-matrix, which
sums up the ladder diagrams and is generally nonlocal.
\begin{figure}[tb]
\centerline{\psfig{file=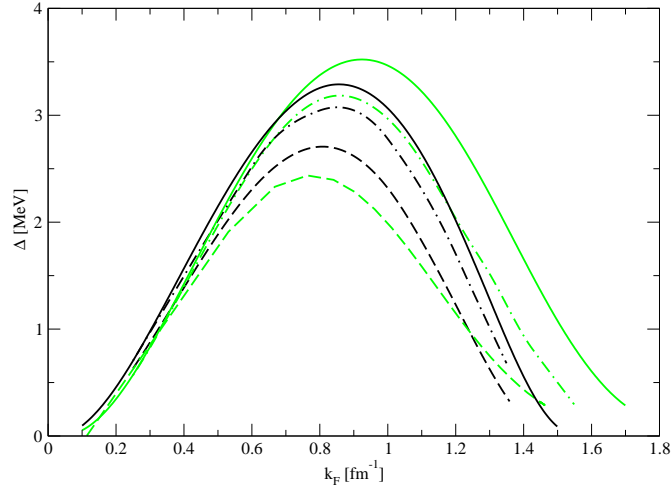,width=3.2in,angle=-90}}
\caption{Singlet $S$-wave ($^1S_0$) pairing gap in neutron matter 
and nuclear matter versus Fermi momentum. The heavy and light 
lines correspond respectively to the phase-shift-equivalent 
Nijmegen interaction and the Gogny effective interaction DS1.  
Solid lines and dashed-dotted lines label results for pure neutron 
matter using the free-space single-particle spectrum and implementing
single-particle renormalization, respectively, while the
dashed line refers to symmetrical nuclear matter with
single-particle renormalization.
\label{fig:NEUTRON_MATTER_M_AND_Z}
}
\end{figure} 
An alternative is to replace the $T$-matrix by an effective 
time-local interaction that is fitted to properties of finite nuclei
(e.g. a Skyrme or Gogny force), in which case Eq.~(\ref{SIGMA}) reduces 
to a mean-field Hartree-Fock approximation.  However, one must be aware 
that the normal-state spectrum itself must depend on the anomalous 
self-energy $\Delta(p)$.  The common replacement of $G(p)$ by $G^N(p)$ 
when computing the normal-state spectrum is an {\it approximation} (sometimes
called the ``decoupling approximation''), which is justified when the
pairing effects can be viewed as a perturbation to the normal state, 
but must be made with care.

Fig.~\ref{fig:NEUTRON_MATTER_M_AND_Z} shows the $^1S_0$ 
pairing gap in neutron matter and symmetrical nuclear matter 
for the high-precision phase-shift-equivalent Nijmegen potential 
and the effective Gogny DS1 force, for different approximations
to the single-particle spectrum.\cite{TOM}  
In the case of smooth effective 
forces such as the the Gogny interaction, a Hartree-Fock approximation 
to the normal self-energy is suitable and was adopted; 
for the (realistic) Nijmegen potential, the $T$-matrix was calculated
in Brueckner theory.  In both cases, the full momentum-dependent
self-energies Re$\Sigma (\vecp)$ were used in the gap equation. The
momentum renormalization yields an effective mass $m^*/m$ less than
unity, thus reducing the density of states $\nu(p_F)$ on the Fermi surface 
and hence also reducing the size of the gap.  The wave-function 
renormalization factor $Z(\vecp)$ is also less than one, leading to 
an additional suppression of the gap.  However, the the magnitude of 
this effect is yet to be established.\cite{BALDO_Z,LOMBARDO_Z}

\subsection{Polarization effects}
\label{Polarization}
An improvement upon the mean-field BCS approximation to fermion 
pairing is achieved in theories that take into account the 
modifications of the pairing interaction due to the background 
medium.  In diagrammatic language, the class of modifications 
known as ``polarization effects'' or ``screening'' arise from the 
particle-hole bubble diagrams, ideally summed to all orders starting 
from the bare interaction as the driving term.  Consider the 
following integral equation describing the four-fermion scattering
process $p_1+p_2 \to p_3+p_4$:\index{polarization effects}
\be \label{PH}
\Gamma (p,p',q) = U(p,p',q)- i\int\!\frac{d^4p''}{2\pi)^4}\,U(p,p'',q)
         G^N(p''+q/2)G^N(p''-q/2)\Gamma(p'',p',q)\,,
\ee
where $q = p_1-p_2$ is the momentum transfer, $p  = p_1 + p_3$, and 
$p' = p_2+p_4$.  Eq.~(\ref{PH}) sums the particle-hole diagrams to 
all orders.  To avoid double summation in the gap equation, the  
driving term $U(p,p',q)$ must be devoid of blocks that contain 
particle-particle ladders.  This driving interaction depends 
in general on the spin and isospin and can be decomposed
as
\be \label{PH2}
U_{\vecq} = f_{\vecq} + g_{\vecq} (\vecsigma \cdot \vecsigma')
          + \left[f'_{\vecq} + g'_{\vecq} (\vecsigma \cdot \vecsigma')
            \right](\vectau\cdot \vectau')\,,
\ee
where $\vecsigma$ and $\vectau$ are the vectors formed from the Pauli
matrices in the spin and isospin spaces.  We assume here that the interaction 
block $U$ depends only on the momentum transfer.  For illustrative 
purposes, the tensor part of the interaction and the spin-orbit terms 
are ignored.  (However, see Subsec.~\ref{ELIASHBERG}, where the tensor 
component is included by means of pion exchange.)  Solution of the 
integral equation (\ref{PH}) then takes the form
\bea 
\nu(p_F)\Gamma_{\vecq} &=& \frac{F_{\vecq}}{1+\Lambda(q)F_{\vecq}}
                   + \frac{G_{\vecq}}{1+\Lambda(q)G_{\vecq}}
                     (\vecsigma \cdot \vecsigma')\nonumber\\
                   &+& \left[\frac{F'_{\vecq}}{1+\Lambda(q)F'_{\vecq}}
                   + \frac{G'_{\vecq}}{1+\Lambda(q)G'_{\vecq}}
                     (\vecsigma \cdot \vecsigma')
\right](\vectau\cdot \vectau')\,,
\eea
where $F_{\vecq} = \nu(p_F)f_{\vecq}$, $G_{\vecq} = \nu(p_F)g_{\vecq}$,
$F'_{\vecq} = \nu(p_F)f'_{\vecq}$, and $G'_{\vecq} = \nu(p_F)g'_{\vecq}$\,, 
while
\be 
\Lambda(q) = \nu(p_F)^{-1}\int\!\frac{d^4p''}{(2\pi)^4}
\,G^N(p''+q/2)G^N(p''-q/2)\,,
\ee
is the (dimensionless) Lindhard function (or polarization tensor).  
\index{polarization tensor}
We are tacitly assuming that the system is in a state characterized
by a well-defined Fermi sphere.  Then, if the momenta of both particles 
lie on the Fermi surface, the momentum transfer is related to the scattering 
angle via $q = 2p_F \sin\theta/2$, and the parameters $F$, $F'$, 
$G$, and $G'$ can be expanded in spherical harmonics with respect to 
the scattering angle, according to
\be \label{LANDAU_EXP}
\left(\begin{array}{c}F(q)\\
                      G(q)\end{array}
\right) =\sum_l \left(\begin{array}{c}F_l\\
                      G_l\end{array}\right) P_l(\cos \theta)\,,
\ee 
and similarly for $F'(q)$ and $G'(q)$.  The Landau parameters 
$F_l$, $G_l$, $F_l'$, and $G_l'$ depend only on the density.
\index{Fermi liquid!parameters}
The isospin degeneracy of neutron matter, reflected in $\vectau \cdot \vectau'
= 1$, implies that the number of independent Landau parameters for each
$\bf q$ or $l$ reduces from four to two, defined by 
$F^{n} = F + F'$ and $G^{n}  = G + G'$.  
Commonly, only the lowest-order harmonics in the expansion 
(\ref{LANDAU_EXP}) are needed.
For a singlet pairing state, in which the total 
spin of the pair is $S = 0$ and
$\vecsigma\cdot\vecsigma' = -3$, the pairing interaction is
given by
\be \label{INDUCED_INT}
\nu(p_F)\Gamma_q = 
F_0^{n}\left[1- \frac{\Lambda(q) F_0^{n}}{1+\Lambda(q)F_0^{n}}\right]
-3G_0^{n}\left[1-\frac{\Lambda(q) G_0^{n}}{1+\Lambda(q)G_0^{n}}\right]\,.
\ee
In general, the polarization tensor $\Lambda(q)$ is complex-valued. 
However, in the limit of zero energy transfer (at fixed momentum), it is 
real and becomes simply
\be 
\Lambda(q)  = -1 + \frac{p_F}{q}\left(1 - \frac{q^2}{4p_F^2}\right)
{\rm ln}\Bigg|\frac{2p_F-q}{2p_F+q}\Bigg|\, 
\ee
in the zero-temperature limit.  Eq.~(\ref{INDUCED_INT}) contains 
two distinct contributions: the direct part generated by the terms 1 
inside the square brackets, and the remaining, induced part 
that accounts for density and spin-density fluctuations.  If the 
Landau parameters are known -- either by inferring them from experiment 
or by computing them within an {\it ab initio} many-body scheme
-- the effect of polarization can be assessed by defining an averaged 
interaction 
\be 
\Gamma = \frac{1}{2p_F^2}\int_0^{2p_F} dq\, q \Gamma (q) \,.
\ee 
The effect of density fluctuations $\propto F^n_0$ is to 
enhance the attraction in the pairing interaction\cite{PETHICK} 
and therefore increase the gap, while the spin-density fluctuations 
$\propto G^n_0$ tend to reduce the attraction and decrease
the gap.\cite{CLARK76}   At densities typical of the inner
crust of a neutron star, the values of microscopically derived 
Landau parameters imply that the suppression of pairing via 
spin-density fluctuations is the dominant effect.\cite{CLARK76} 
The Landau parameters of neutron matter and symmetrical nuclear
matter have been studied extensively within complex many-body 
schemes\cite{LP1,LP2,LP3} whose description is beyond
the scope of this chapter.  It should be noted, however, that the 
matrix elements of the pairing interactions derived in some of
these schemes, including the Babu-Brown approach\cite{LP1,LP2} 
and its extensions such as polarization-potential theory, 
\index{polarization potential theory}
\cite{AINSWORTH,WAMBACH} have 
been used in conjunction with the weak-coupling approximation 
(\ref{WCF}), which -- as indicated above -- is generally 
inadequate in the nuclear/neutron-matter context.
\begin{figure}[bt]
\psfig{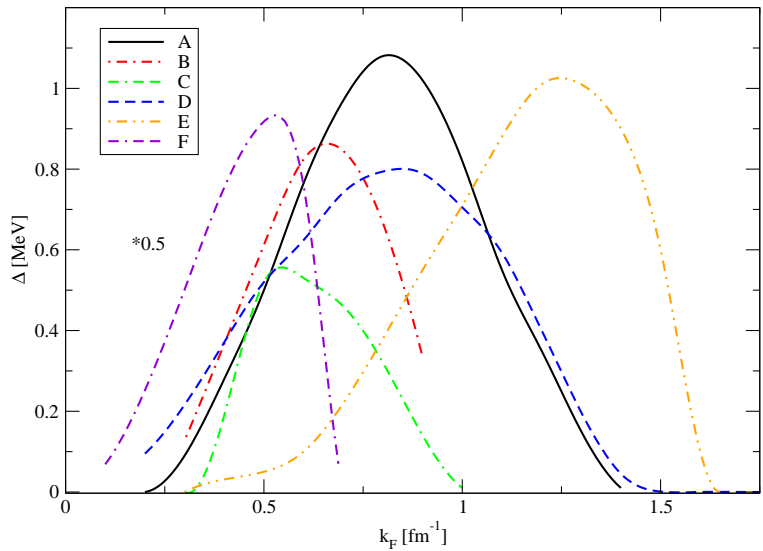}
\vskip 0.2 cm
\caption{
Singlet $S$-wave ($^1S_0$) pairing gaps in neutron matter versus 
Fermi momentum, as obtained from microscopic calculations that attempt to 
account for medium-modification of the pairing interaction
and self-energies.  The curves are labeled as:  
A - Wambach et al. [42], B - Chen et al.  [48], 
C - Chen et al. [49], 
D - Schwenk et al. [52], 
E - Schulze et al.  [51],
F - Fabrocini et al.  [50].
\label{fig:screening}
}
\end{figure}

The initial microscopic calculations of the effects of medium
polarization on pairing were carried out within an alternative 
many-body approach, the method of Correlated Basis Functions 
\index{correlated basis functions (CBF)}
(CBF)\cite{CBF1,CBF2,CBF3,CBF4,CBF5,CHEN86,CHEN93} (to be 
described in Subsec.~\ref{sec:CBF}).  
At a qualitative level, the findings of the CBF studies of
Chen et al.\cite{CHEN86,CHEN93} are consistent with much of the
later work based on Green's functions and Fermi-liquid theories.  
Until recently, there was broad agreement that the screening 
reduces the singlet $S$-wave gap by factor of 3 or so 
(c.f.~Fig.~\ref{fig:NEUTRON_MATTER_M_AND_Z}).  However, the 
density profiles of the calculated gaps differ considerably.  
This is illustrated in Fig.~\ref{fig:screening}, which presents
a composite plot of theoretical predictions for the dependence
of the $^1S_0$ pairing gap at the Fermi surface, 
$\Delta(k_F)$, upon the Fermi wave number $k_F=p_F/\hbar= (3\pi^2\rho)^{1/3}$. 
The six curves in the plot correspond to various microscopic approaches
that include a screening correction. The results of 
Wambach et al.,\cite{WAMBACH} Schulze et al.,\cite{SCHULZE96} and
Schwenk et al.\cite{SCHWENK} are based on microscopic treatments
rooted in Landau/Fermi-liquid theory, with polarization-type diagrams
summed to all orders.  The results of Chen et al.\cite{CHEN86,CHEN93}
and Fabrocini et al.\cite{FABROCINI} were obtained within two different
implementations of CBF theory.  

Further assessment of the status of quantitative microscopic
evaluation of the singlet-$S$ gap is deferred until 
Subsec.~\ref{sec:CBF}, where the elements of CBF 
approaches to the pairing problem are reviewed.  We may 
already remark, however, that explicit comparison of the
pairing matrices constructed in the different theories
could help to eliminate discrepancies introduced by use
of the weak-coupling approximation in some of the theoretical
treatments.  Another important consideration is consistent
inclusion of medium effects on both the pairing interaction
and the self-energies.

\subsection{Non-adiabatic superconductivity}
\label{ELIASHBERG}
Since mesons propagate in nuclear matter at finite speed, the  
interactions among nucleons are necessarily retarded in character.  
As a consequence, the self-energies (and in particular the gap function) 
must depend on energy or frequency. Within the meson-exchange picture 
of nuclear interactions, the lightest mesons -- pions -- should 
be the main source of nonlocality in time.  This suggests that
it may be fruitful to consider a pairing model in which the 
interactions are modeled in terms of pion exchanges, plus 
contact terms that can be approximated by Landau parameters.
\index{Fermi liquid}
\index{Fermi liquid!parameters}
Thus one assumes an interaction structure
\index{non-adiabatic superconductivity}
\bea \label{PION_INTERACTION}
V_{NN} = -\frac{f_{\pi}}{m_{\pi}} 
(\vecsigma \cdot \vecnabla ) (\vectau \cdot \vecphi )
+\tilde U(q)
\eea
in which $ \vecphi$ is the pseudoscalar isovector pion field satisfying
the Klein-Gordon equation, $f_{\pi}$ is the pion-nucleon coupling 
constant, and $m_{\pi}$ is the pion mass.  
Here $\vecphi$ is the pseudoscalar isovector pion field satisfying
the Klein-Gordon equation, $f_{\pi}$ is the pion-nucleon coupling 
constant, and $m_{\pi}$ is the pion mass. The operator structure of 
the term $\tilde U(q)$ is like that of Eq.~(\ref{PH2}),
but with constants differing numerically since the tensor 
one-pion exchange is treated separately.  

For {\it static pions}, the one-pion-exchange two-nucleon
interaction in momentum space is given by
\be\label{ope} 
V_{\pi}(\vecq) = -\frac{f_{\pi}^2}{3m_{\pi}^2}\, 
\frac{\vecq^2}{\vecq^2+m_{\pi}^2}
\left[\vecsigma _1 \cdot \vecsigma _2 +S_{12}(\vecn)\right]
\vectau_1 \cdot \vectau_2\,,
\ee
where $\vecq$ is the momentum transfer, $\vecn =\vecq /q$, 
and $S_{12}(\vecn)$  is the tensor operator.  This interaction
is known to reproduce the low-energy phase shifts and, to a large
extent, the deuteron properties.\cite{WEISE}  Below, however, the pairing
correlations are evaluated from the diagrams that contain {\it
dynamical pions}, with full account of the frequency dependence of the 
pion propagators.  The static results can be recovered, and a 
relation to the phase shifts established, only in the limit 
$\omega \to 0$ in the pion propagator.  The dynamics at intermediate
and short range is dominated in turn by the correlated two-pion,
$\rho$-meson, and heavier-meson exchanges.  Short-range
correlations are crucial for a realistic description of low-energy
phenomena, being necessary to achieve nuclear saturation.
Moreover, the response functions calculated from one-pion exchange 
alone would already precipitate a pion-condensation instability
in nuclear matter at an unrealistically low density.

The interaction (\ref{PION_INTERACTION}) leads to a time-nonlocal 
formulation of nuclear superconductivity in neutron-star 
matter.\cite{SEDRAKIAN03}  The Dyson-Schwinger equation 
\index{Dyson-Schwinger equations} 
(\ref{DYSON}) provides the starting point.  Since the formulation
will incorporate the full frequency dependence of the self-energies, 
it is useful (and also conventional~\cite{ELIASHBERG}) 
to define the wave-function 
renormalization differently than in Subsec.~\ref{BCS}.  Thus
we set $Z(p) = 1-\omega^{-1}\Sigma_A(p)$, the retarded self-energy 
being decomposed into components even ($S$) and odd ($A$) 
in $\omega$, i.e.\ $\Sigma(p) = \Sigma_S(p) + \Sigma_A(p)$.  
The single-particle  energy is then renormalized as 
$E_S= \epsilon_p+\Sigma_S(E_S,\vecp )$.
Accordingly, the propagators now take the forms
\begin{figure}[t]
\centerline{\psfig{file=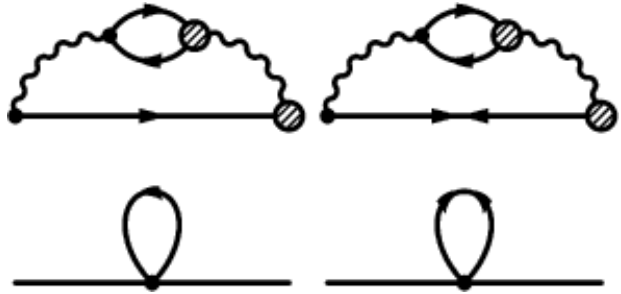,width=3.2in,angle=0}}
\caption{{\it Top panel}.  Baryon Fock-exchange self-energies 
for normal (left graph) and anomalous (right graph) sectors.
The solid lines correspond to fermions; and the wavy lines, 
to pions.  The blobs symbolize RPA-renormalized vertices; 
and the dots, bare pion-nucleon vertices. 
{\it Bottom panel}.  Normal and anomalous Hartree diagrams. 
The dots stand for contact Landau interactions; the straight
lines are shown for clarity.\label{HF_DIAGRAM}
}
\end{figure}
\bea\label{PROP1}
G(p) &=& \frac{\omega Z(p)+E_S(p)}
{(\omega+i\eta)^2 Z(p)^2-E_S(p)^2-\Delta(p)^2}\,,\\
F(p) &=&- \frac{\Delta(p)}
{(\omega+i\eta)^2 Z(p)^2-E_S(p)^{2}-\Delta(p)^2}\,,
\eea
where $\Delta\Delta^{\dagger}\equiv -\Delta^2$.  The  
self-energies of the theory are shown in Fig.~\ref{HF_DIAGRAM}. 
The analytical counterparts of the Fock self-energies are 
\bea \label{24}
\Sigma^{\rm Fock}(\omega,\vecp)  &=&-2\int\!\!\frac{d^3q}{(2\pi)^3} 
\int_{-\infty}^{\infty}\!\!\frac{d\ep}{2\pi}\Xi_{0}(\vecq)
{\rm Im}G(\ep ,\vecp-\vecq)C(\omega , \ep , \vecq) \Xi (\vecq)\,,\\
\label{25}
\Delta^{\rm Fock}(\omega,\vecp)  &=&-2\int\!\!\frac{d^3q}{(2\pi)^3} 
\int_{-\infty}^{\infty}\!\!\frac{d\ep}{2\pi}
\Xi_{0}(\vecq){\rm Im}F(\ep ,\vecp-\vecq)
C(\omega ,\ep , \vecq)\Xi(\vecq)\,,
\eea
where
\be \label{26}
C(\omega , \ep , \vecq) = 
\int_{0}^{\infty}\!\!\frac{d\omega '}{2\pi}B(\omega' , \vecq)
\left[
\frac{f(\ep)+g(\omega')}{\ep-\omega'-\omega-i\eta}
+\frac{1-f(\ep)+g(\omega')}{\ep+\omega'-\omega-i\eta}
\right]\,.
\ee
Here $B(q)$ is the pion spectral function, while $\Xi^0(q)$ and $\Xi(q)$
are the bare and renormalized pion-neutron vertices. One remarkable 
feature of Eqs.~(\ref{24})--(\ref{26}) is that the energy and momentum 
dependence of the self-energies is determined by the dynamical features 
of the meson (here pion) field.  Another salient feature is that 
the normal and anomalous sectors are coupled, in contrast to the BCS case,
where the unpaired single-particle energy is unaffected by the pairing.
The most important contribution to the pion spectral function comes 
from the coupling to virtual particle-hole states, which are described 
by the (retarded) particle-hole polarization tensor $\Pi(\omega,\vecp)$. 
Specifically, one finds \index{spectral function!pions}
\be 
B(q) = \frac{-2{\rm Im}\Pi^R(q)}
{[\omega^2-\vecq^2-m_{\pi}^2-{\rm Re}\Pi^R(q)]^2
+[{\rm Im}\Pi^R(q)]^2}\,.
\ee
\begin{figure}[tb]
\centerline{\psfig{file=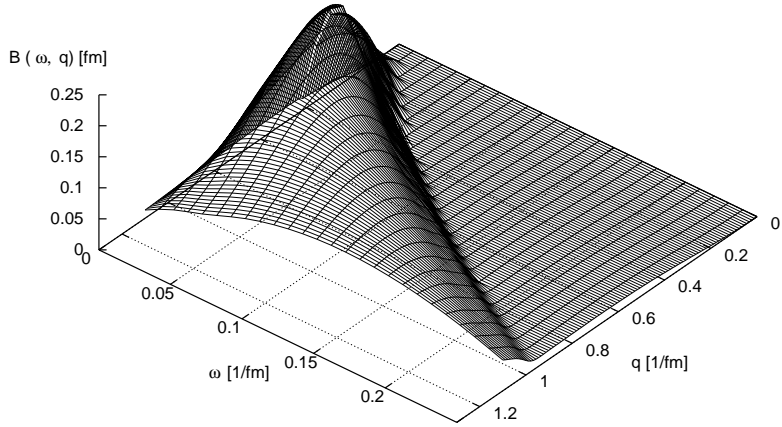,width=3.2in,angle=-90}}
\caption{Pion spectral function in neutron matter 
as a function of energy and momentum transfer. 
The density corresponds to $k_F=0.55$ fm$^{-1}$.
}\label{fig:PION_SPEC}
\end{figure}
The spectral function of pions in neutron matter is illustrated in 
Fig.~\ref{fig:PION_SPEC} for the momentum transfer 
range $0\le q\le 2k_F$, where $k_F = 0.55$ fm$^{-1}$. It is seen 
that (i)~the spectral function has a substantial weight for finite 
energy transfer, the maximum being determined by the pion dispersion 
relation $\tilde \omega^2  =\vecq^2 + m_{\pi}^2  
+ {\rm Re}\Pi(\tilde\omega,q)$ and (ii) the
spectral function is substantially broadened due to the excitations 
of particle-hole pairs, which are treated in the random-phase 
approximation (RPA).\cite{WEISE} \index{random-phase approximation}
In addition to the Fock self-energies we need to include the 
Hartree contribution, which reduces to 
\bea 
\Delta^{\rm BCS}(\omega,\vecp) &=& -2(F^n_0-3G^n_0)
\int\frac{d^4p'}{(2\pi)^4}
{\rm Im}\,F  (\omega+\omega',\vecp+\vecp')\,.
\eea
\begin{figure}[tb]
\centerline{\psfig{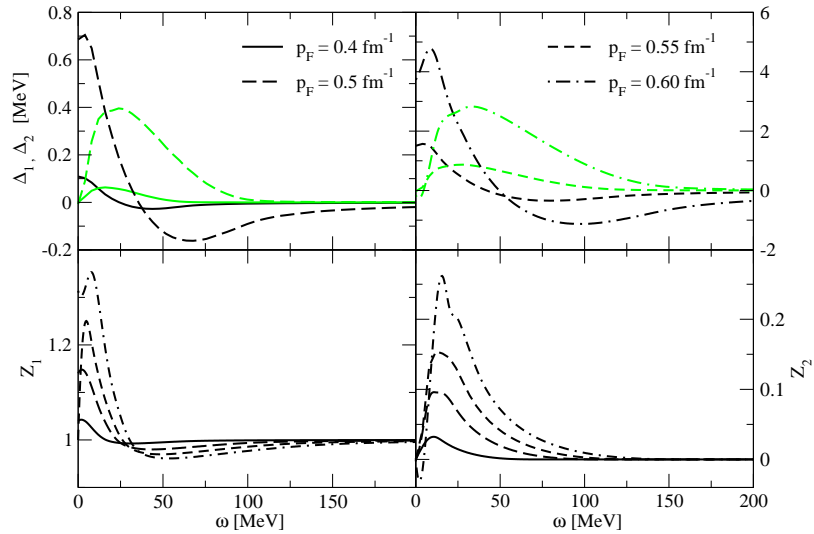}}
\caption{{\it Top panel.}  Frequency dependence of real (heavy lines) 
and imaginary (light lines) parts of the gap function, $\Delta_1(\omega)$   
and $\Delta_2(\omega)$, respectively, for $k_F=0.4$ (solid), 0.5 
(long-dashed), 0.55 (dashed), and 0.6 (dashed-dotted lines) fm$^{-1}$. 
The on-shell values of the pairing gap are 0.1, 0.7, 1.4, and 3.7 MeV 
for $k_F=0.4$, 0.5, 0.55, and 0.6 fm$^{-1}$. 
{\it Bottom panel.} Frequency dependence of real 
(left panel) and imaginary (right panel) parts of the wave-function 
renormalization, $Z_1(\omega)$ and  $Z_2(\omega)$.  Labeling is the 
same as in the top panel. \label{fig:NONLOCAL_GAPS}
}
\end{figure}
Solutions of the self-consistent equations (\ref{25}) and (\ref{26})
are shown in Fig.~\ref{fig:NONLOCAL_GAPS} at zero temperature and 
for densities specified by the indicated Fermi momenta.  At small 
energy transfers, the imaginary components of the gap and 
wave-function renormalization vanish, and one recovers the BCS 
limit.  For finite energy transfers these functions develop substantial 
structure that reflects the features of the pion spectral function
[the driving term in the kernel of integral equations for 
$\Delta(q)$ and $Z(q)$].  Note that the actual value of the gap 
on the mass shell does depend on the detailed structure of these 
functions far from the mass shell.  However, it is possible 
to renormalize the pion spectral function such that the high-energy
tails are eliminated while the on-mass-shell physics 
is unchanged.

Theories that explicitly include the light mesons -- pions or 
kaons -- in the computational scheme have the advantage that  
they embody the precursor phenomena associated with the softening of 
the pion (kaon) modes close to the threshold for condensation. 
In the case of $P$-wave pairing, an enhancement of the pairing 
correlation has been predicted.\cite{KHODEL}

\subsection{The method of correlated basis functions (CBF)}
\label{sec:CBF}
The variational formulation of BCS theory is based on 
the trial wave function (BCS state)
\index{correlated basis functions (CBF)}\index{wave function!trial}
\be\label{BCS_WF} 
\Psi_{\rm BCS} = \prod_{\vecp} 
\left[(1-h_{\vecp})^{1/2}+h_{\vecp}^{1/2}
\psi^{\dagger}_{{\vecp}\uparrow}\psi^{\dagger}_{-{\vecp}\downarrow}\right]
\vert 0\rangle\,,
\ee 
where the real function $h_{\vecp}$, which gives the occupation
probability of the pair state $({\vecp}\uparrow , -{\vecp}\downarrow)$,
is subject to variation.  For this trial state one may compute the 
anomalous density 
\be \label{CHI}
\chi_{\vecp} = \langle\Psi_{\rm BCS} \vert 
\psi^{\dagger}_{{\vecp}\uparrow}\psi^{\dagger}_{-{\vecp}\downarrow}
\vert \Psi_{\rm BCS}\rangle = h_{\vecp}^{1/2}(1-h_{\vecp})^{1/2}\,,
\ee
whose Fourier image specifies the spatial structure of a Cooper 
pair.  The standard coupled BCS equations for the energy gap, the
fermionic density, and the quasiparticle energy at zero 
temperature, namely
\be
\Delta(p) =-\int \frac{d^3 p'}{(2\pi)^3}V( p, p')\frac{1}{2E(p')}
\Delta( p')\,, \label{bcs0}
\ee
\be
\rho = \int {\frac{d^3 p}{(2\pi)^3}}h_{\vecp} =\int\frac{p^2 dp}{(2\pi)^2}
       \left[1 - \frac{\varepsilon(p) - \mu }{2E(p)}\right] \,,  
\label{dens}
\ee
\be
E(p) = \sqrt{ \left[\varepsilon(p) - \mu\right]^2 + \Delta^2(p) } \,,
\label{qp}
\ee
are generated naturally upon (i) evaluating 
the expectation value of the grand-canonical Hamiltonian for the BCS 
state (\ref{BCS_WF}), and (ii) performing a variational minimization 
of this functional with respect to $h_{\vecp}$, under the constraint 
that the expectation value of the particle-number density 
coincides with the prescribed density.

Treatment of strongly interacting fermionic systems (including
nuclear problems) within a variational framework calls for 
improved superfluid trial states.  The systems of interest
are characterized by a bare two-body interaction containing
a strong inner repulsive core along with longer-range
attractive components.  To obtain a reasonable energy 
expectation value, the trial function must adequately describe
the short-range geometric correlations induced by the
repulsion, which inhibits the close approach of a 
pair of particles.\cite{CBF1,CBF2,CBF3,CBF4}
The simplest choice involves the Jastrow correlation factor
\index{Jastrow correlation factor}
\be \label{JASTROW}
F_J = \prod_{i< j}^N f(r_{ij})\,, \quad 
\mathop {\lim }\limits_{r \to 0}f(r)\to 0\,,\quad
\mathop {\lim }\limits_{r \to \infty }f(r)\to 1\,,
\ee
which is suited to efficient description of {\it state-independent} 
two-body correlations, especially the short-range repulsive
effects.  As a bonus, with proper optimization of 
the two-body function $f(r)$, the Jastrow factor can also 
incorporate effects of virtual phonon excitations and indeed
can reproduce the correct asymptotic behavior of long-range 
correlations.

A substantially improved trial superfluid state of definite particle 
number may be formed by applying the Jastrow operator (\ref{JASTROW}) 
to an $N$-particle projection of the BCS trial state, expressed 
in the configuration-space representation as
\be 
\Phi^{(N)}_{\rm BCS} = \frac{1}{\sqrt{\cal N}}\,{\cal A}
\{\phi(\vecr_1,\vecr_2),\phi(\vecr_2,\vecr_3)\dots 
\phi(\vecr_{N-1},\vecr_N)\}\,.
\label{proj}
\ee
Here,
$\phi({\vecr_i,\vecr_j})$ is the antisymmetrized Fourier 
image of $\chi_{\vecp}$ given by Eq.~(\ref{CHI}), times a 
spin function, ${\cal A}$ is an antisymmetrization operator 
acting on particle pairs in different $\phi$ factors, and 
${\cal N}$ is a normalization constant.   In spirit, this
is the approach adopted in the early work of Yang and 
Clark,\cite{YANG1,YANG2} although in practice they determined the 
one-body, two-body, etc.\ density matrices required for their 
cluster-expansion treatment of the short-range correlations 
{\it from the BCS state} (\ref{BCS_WF}), rather than from 
its $N$-particle projection (\ref{proj}).

Following up on the work of Refs.~\refcite{YANG1,YANG2}, a formal 
variational theory\cite{KRO_CLARKIII} \index{variational theories}
of the superfluid ground state 
of uniform, infinite nucleonic systems was developed for a trial 
correlated BCS state constructed in Fock space,
\be
|\Psi_s \rangle = \sum_N \sum_{m^{(N)}} \hat F^{(N)}|\Phi_m^{(N)} \rangle
 \langle \Phi_m^{(N)} | {\rm BCS} \rangle \,,
\label{CBCS}
\ee
where $\hat F^{(N)}$ is an unspecified correlation operator
meeting certain minimal conditions and
$\{|\Phi_m^{(N)}\rangle \}$ is a complete set of Fermi-gas
Slater determinants, both referred to the $N$-particle
Hilbert space.  Thus, the correlated normal states 
$\{\hat F^{(N)} |\Phi_m^{(N)}\rangle \}$ are superposed with
the same amplitudes as the model states $|\Phi_m^{(N)}\rangle $ 
have in the corresponding grand-canonical representation of 
the original BCS state (\ref{BCS_WF}).  Repetition of steps 
(i) and (ii) above for this correlated superfluid {\it Ansatz} 
yields a theory having the same structure as ordinary BCS 
theory, when the ``decoupling approximation'' is applied.  In 
this approximation, only one Cooper pair at a time is considered,
while treating the background as normal.  Formally, 
the expectation value of the grand-canonical Hamiltonian is 
expanded in terms of the deviations of the Bogolyubov amplitudes 
$u_{\vecp} = (1 - h_{\vecp})^{1/2}$ and $v_{\vecp} = h_{\vecp}^{1/2}$ 
about their normal-state values, retaining deviant terms at most 
of first order in $v_{\vecp}^2 - \theta(p)$ and second order 
in $u_{\vecp}v_{\vecp}$, where $\theta(p)$ is the Fermi step.

Within this framework, the gap equation and density constraint
maintain the same mean-field forms as obtained for the bare
BCS state, except for the attachment of renormalization factors 
$z_p^{-1}$ to the gap function $\Delta(p)$ when it appears
in quasiparticle energy denominators.  The presence of
correlations introduced by the operator ${\hat F}^{(N)}$ is 
otherwise reflected only in the replacement of the pairing matrix 
elements $V(p,p')$ and single-particle energies $\varepsilon(p)$ 
derived from the bare interaction based on the BCS trial state 
(\ref{BCS_WF}) and variational steps (i)-(ii), by {\it effective} pairing 
matrix elements ${\cal V}(p,p')$ and {\it correlation-dressed} 
single-particle energies $\xi(p)$ built from combinations 
of diagonal and off-diagonal matrix elements of the Hamiltonian
and unit operators in the correlated normal bases
$\{ {\hat F}^{(N)}|\Phi_m^{(N)} \rangle \}$.  When the 
state-independent Jastrow choice $F_J$ of Eq.~(\ref{JASTROW}) 
is assumed for the correlation operator $\hat F$, the dressed
quantities ${\cal V}(p,p')$ and $\xi(p)$ can be evaluated by Fermi 
hypernetted-chain (FHNC) \index{hypernetted chain methods}
methods developed in Ref.~\refcite{KRO_CLARKII},
with results for neutron matter and liquid $^3$He reported
in Ref.~\refcite{KRO_CLARKIII}. 

An important advance in the CBF approach to pairing was made
in Ref.~\refcite{CBF5}, where the variational
description was extended to create a correlated-basis
perturbation theory for the exact superfluid ground state.  
Again imposing the decoupling approximation, a sequence of
approximations to the grand-canonical energy may be defined, 
each preserving under variation the standard form of the 
gap equation, but with successive 
improvements on the effective pairing matrix elements 
and dressed self-energies.  (A convenient modification
of the trial correlated ground state (\ref{CBCS}) was made 
by inserting the normalization factor $\langle \Phi_m^{(N)} 
|{ F^{(N)}}^\dagger F^{(N)} | \Phi_m^{(N)} \rangle^{-1/2}$ inside the 
summations.  This eliminates the renormalization factors $z_p^{-1}$ 
mentioned above.)  Making the Jastrow choice for the correlation
operator $\hat F$, the leading perturbative corrections
to the variational results for ${\cal V}(p,p')$ and $\xi(p)$
were generated, represented in diagrammatic form, and evaluated.
These corrections include the leading contribution from medium polarization
\index{polarization effects}
within the CBF framework.  Here we should point out that the terms 
in the CBF perturbation expansions of the various quantities 
are not easily interpreted in terms of conventional Goldstone or
Feynman diagrams, although they may have similar appearance.  
A given order in the CBF expansion, including the ``zeroth-order'' 
variational term, will contain pieces belonging to
any number of perturbative orders in the conventional 
sense.  In general there will be terms accounting for the 
nonorthogonality of the basis, terms that correct the 
average-propagator approximation inherent (for example) 
in the Jastrow description of $\hat F$-correlations, terms that 
correct for non-optimality of the chosen $\hat F$-correlations,
etc.

In microscopic studies of nucleonic systems, it is generally
imperative to include the effects of state-dependent correlations 
arising from realistic $NN$ interactions which contain,
separately in each spin and isospin channel, contributions
of central, tensor, and spin-orbit character.  In principle,
the CBF perturbation expansions provide for systematic
correction of the Jastrow {\it Ansatz} for the correlation
operator $\hat F$, so as to incorporate these state-dependent 
effects.  However, it is clearly preferable to take account
of state dependence already in the choice of $\hat F$,
thus reducing the need to correct the variational treatment
with CBF perturbation theory.  A suitably general correlation 
{\it Ansatz}, within the class containing only two-body
correlation factors, is given by
\be
{\hat F} = \prod_{i<j}^N f(ij) \,, \qquad f(ij) = 
\sum_{\alpha=1}^n f_\alpha(r_{ij}) o_\alpha(ij) \,,
\label{sdJASTROW}
\ee
where $f(ij)$ contains terms for the same operators $o_\alpha(ij)$
as are present in the assumed realistic $NN$ interaction (e.g., 
the Argonne $v_{18}$ model\cite{WIRINGA1}), or an adequate subset of 
them.  Profound difficulties arise in the implementation of this 
choice, due to non-commutativity among the $o_\alpha(ij)$
operators.  The analog of FHNC resummation being still beyond
our reach for such state-dependent correlations, existing 
calculations proceed with straightforward cluster or 
power-series expansions, perhaps with vertex 
corrections.\cite{CHEN93,FABROCINI}
It is important to appreciate that the extended Jastrow form 
(\ref{sdJASTROW}) of the $\hat F$-operator is equipped to include the
lion's share of the polarization corrections (just as the
simple state-independent Jastrow form is capable of capturing
the major effects of density-density fluctuations).

Chen et al.\cite{CHEN86} applied CBF pairing theory as developed 
in Ref.~\refcite{CBF5} to superfluid neutron matter 
in the $^1S_0$ phase, assuming state-independent Jastrow correlations 
and taking account of the leading CBF perturbation corrections 
to the variational treatment.  The polarization and other 
corrections produced a very substantial suppression from the 
variational estimate of the gap $\Delta(k_F)$, the peak 
value being reduced by a factor $\sim 4$ and situated at lower density.  
This treatment was updated in Ref.~\refcite{CHEN93}.  Major
improvements were made in the choice of the variational two-body 
correlation functions.  Significantly larger gap values were 
obtained, but again the perturbative corrections were estimated 
to suppress the peak value by a factor $\sim 4$ and shift its location 
to a lower density.  Quantitatively, the results of the later of 
the two perturbative CBF calculations are the more reliable.
  
Shortly after the work of Krotscheck and Clark,\cite{KRO_CLARKIII}
an independent approach to CBF description of pairing based on the trial
superfluid state (\ref{CBCS}) was launched by Fantoni.\cite{FANTONI81} 
With immediate specialization of the correlation operator $\hat F$
to the state-independent Jastrow form (\ref{JASTROW}), it proved
feasible to extend the diagrammatic techniques of standard
Fermi hypernetted-chain theory\cite{CBF2,FANTONI_ROSATI} and thereby
\index{hypernetted chain methods}
enable practical evaluation of the one-body density matrix and 
radial distribution function associated with the correlated BCS state 
(\ref{CBCS}).  Derivation of the corresponding gap equation and 
density constraint was achieved without resorting to the decoupling 
approximation.  

Quite recently,\cite{FABROCINI} Fantoni's CBF 
approach has been generalized -- insofar as practicable -- to 
include state-dependent correlations of the form (\ref{sdJASTROW}).  
The results for the $^1S_0$ neutron gap lie distinctly higher than the
results of earlier work designed to include nontrivial medium effects
on pairing (see Fig.~\ref{fig:screening}).  Thus, they conflict with
the general consensus that these effects lead, on balance, to a strong 
suppression of the gap value.  Concurrent estimates\cite{FABROCINI} of the 
gap based on an auxiliary-field diffusion Monte Carlo (AFDMC) calculation 
\index{Monte-Carlo methods}
lie even higher than the new CBF estimates (by roughly 0.5 MeV, with 
a peak value of more than 2.5 MeV at $k_F \simeq 0.6\,{\rm fm}^{-1}$).  
A recent numerical study of $^1S_0$ pairing in neutron matter within the 
self-consistent Green's function (SCGF) method\cite{SDPAIRING4}
gives results in essential agreement with the AFDMC estimates.
One might conclude from this agreement that the medium-polarization
effects, arising from the exchange of spin-density fluctuations and 
other virtual processes,\cite{CLARK76} are less important than 
previously imagined.  On the other hand, the SCGF calculation,
by construction, neglects such collective correlations of longer 
range, while the AFDMC stochastic estimates, obtained for relatively
small samples of neutrons, might also fall short in their inclusion
of these effects.  At any rate, the latest computational results 
continue to highlight the extreme sensitivity of the $^1S_0$ 
pairing gap to the assumptions made in pursuing its evaluation by 
microscopic methods.  The quantitative situation for pairing in 
spin-triplet $T=1$ states is even less clear.\cite{KHODEL,FRIMAN}

\section{Pairing in asymmetric nuclear systems}
\label{SECTION3}

\index{nuclear matter!asymmetric}
The isospin asymmetries characteristic of neutron-star cores, 
with proton fractions $\sim 5\%$, are too large to permit 
isospin-singlet (neutron-proton) pairing.  
A possible exception involves Bose-Einstein condensation of kaons
\index{kaon condensation}
at densities several times nuclear saturation density, in which 
case the matter is approximately isospin-symmetric.  
In high-density isospin-symmetric nuclear matter, neutron-proton pairs 
form in the $^3D_2$ partial wave.\cite{D_WAVE1,D_WAVE2}  However, 
once the isospin symmetry is slightly broken, $^3D_2$ pairing is 
suppressed and isospin-triplet neutron-neutron and proton-proton 
pairs are formed.  Due to their large partial density, neutrons pair 
in the $^3P_2$--$^3F_2$ tensor-coupled channel,\cite{P_WAVE0,P_WAVE} 
while the less abundant protons pair in the $^1S_0$ 
state.\cite{CHAO,TAKATSUKA81}

We have seen in Subsec.~(\ref{GFA}) that for fermionic systems 
which are invariant under reversal of time and reflections 
of space, the quasiparticle spectrum is symmetric under $p\to-p$, and 
consequently the antisymmetric piece $E_A$ in 
Eq.~(\ref{BRANCHES}) vanishes.  Depending on the system, these symmetries 
could be broken either by the presence of external gauge fields 
or due to intrinsic properties such as the mass difference in 
mixtures of gases. At any rate, we now focus on systems having 
$E_A\neq 0$, and the pairing in question is between fermions 
that lie on different Fermi surfaces.  We shall call such 
systems {\sl asymmetric superconductors}(hereafter ASC).

Initial studies of ASC where carried out in the early sixties when, 
shortly after development of the BCS theory of superconductivity, 
metallic superconductors with paramagnetic impurities were 
studied experimentally.\cite{CLOGSTON,CHANDRA,SARMA,GR}  
Since collisions with impurities can flip the spins of electrons,
an imbalance between spin-up and spin-down electron populations
is created.  This effect can be mimicked by 
introducing an average, effective magnetic field that lifts the 
electron-spin degeneracy due to its interaction with the electron 
magnetic moment.  The novel aspect of the studies of ASC (apart 
from the new context) is the realization that a correct 
interpretation of the results requires a self-consistent 
solution of the gap and density equations, even in the 
weak-coupling limit where the changes in the value of the 
chemical potential due to pairing are small. 

To avoid undue complications in describing ASC, we will operate
within the framework of conventional BCS theory, in the sense
that effects of wave-function renormalization and medium 
polarization are neglected.  We shall also suppress the additional
complication of the $^3S_1$--$^3D_1$ tensor coupling. This aspect 
is not essential for the present discussion; see 
Refs.~\refcite{SDPAIRING1,SDPAIRING2,SDPAIRING3} for the 
relevant details.

Thus, the equations underlying the theory of ASC are taken to be 
(\ref{GAP_PARTIAL}) and (\ref{DENS1}), the spectrum being given by 
Eq.~(\ref{BRANCHES}) with $E_A \neq 0$.
If the spatial symmetries are unbroken, then 
$E_A=\delta\mu = (\mu_n-\mu_p)/2$, where $\mu_n$ and $\mu_p$ are 
the neutron and proton chemical potentials. In general, 
Eqs.~(\ref{GAP_PARTIAL}) and (\ref{DENS1}) must be solved 
self-consistently.  Consider first the procedure in which 

Eq.~(\ref{GAP_PARTIAL}) is solved by parametrizing the 
asymmetry in terms of the difference in the chemical 
potentials, and the densities of the species are computed 
after the gap equation is solved.  Such an analysis 
predicts\cite{CLOGSTON,CHANDRA,SARMA} a double-valued character of 
the gap as a function of $\delta\mu$.  On the first branch,
the gap has a constant value $\Delta (\delta\mu) = \Delta (0)$ over the 
asymmetry range $0\le\delta\mu \le \Delta(0)$ and vanishes beyond the point 
$\delta\mu = \Delta(0)$.  The second branch exists in the range  
$\Delta(0)/2\le\delta\mu\le\Delta(0)$, with the gap increasing from zero 
at the lower limit to $\Delta(0)$ at the upper limit.  Only the portion
$\delta\mu\le\Delta(0)/\sqrt{2}$ of the upper branch is 
stable, i.e., it is only in this range of asymmetries that
the superconducting state lowers the grand thermodynamic potential
from that of the normal state.\cite{SARMA}  Thus, the dependence of the 
superconducting state on the shift in the Fermi surfaces is 
characterized by a constant value of the gap, which vanishes at the
Chandrasekhar-Clogston\cite{CLOGSTON,CHANDRA} limit 
$\delta\mu_1 =\Delta(0)/\sqrt{2}$. 

A different picture emerges from an alternative treatment of
the problem in which particle-number conservation 
is incorporated explicitly by solving 
Eqs.~(\ref{GAP_PARTIAL}) and (\ref{DENS1}) self-consistently.\cite{SAL,SL00} 
These studies find a single-valued gap 
as a function of the isospin asymmetry 
$\alpha = (\rho_{n}-\rho_{p})/(\rho_{n}+\rho_{p})$.
\begin{figure}[t]
\centerline{\psfig{file=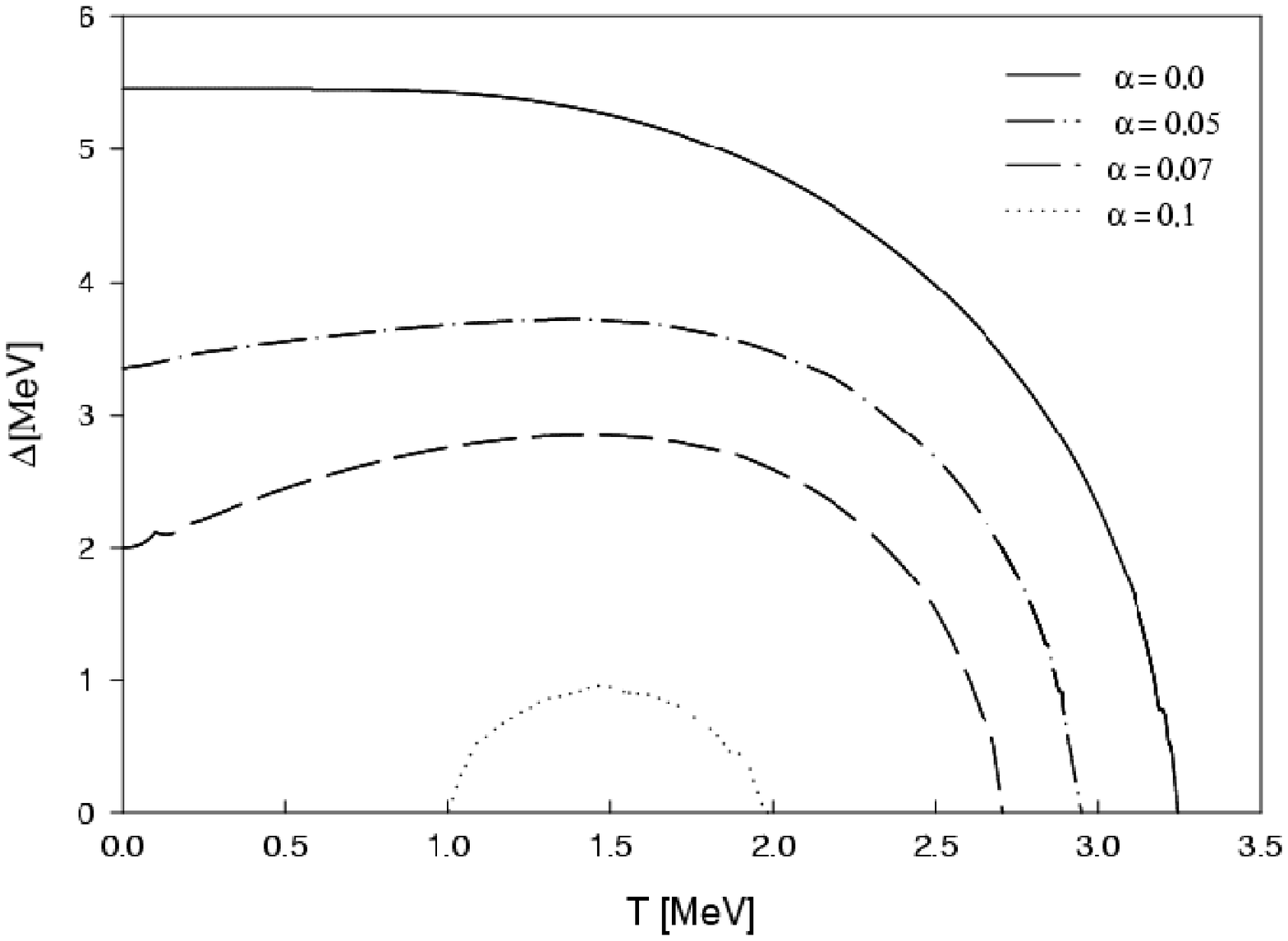,width=2.2in,angle=0}
\hspace{0.2cm}\psfig{file=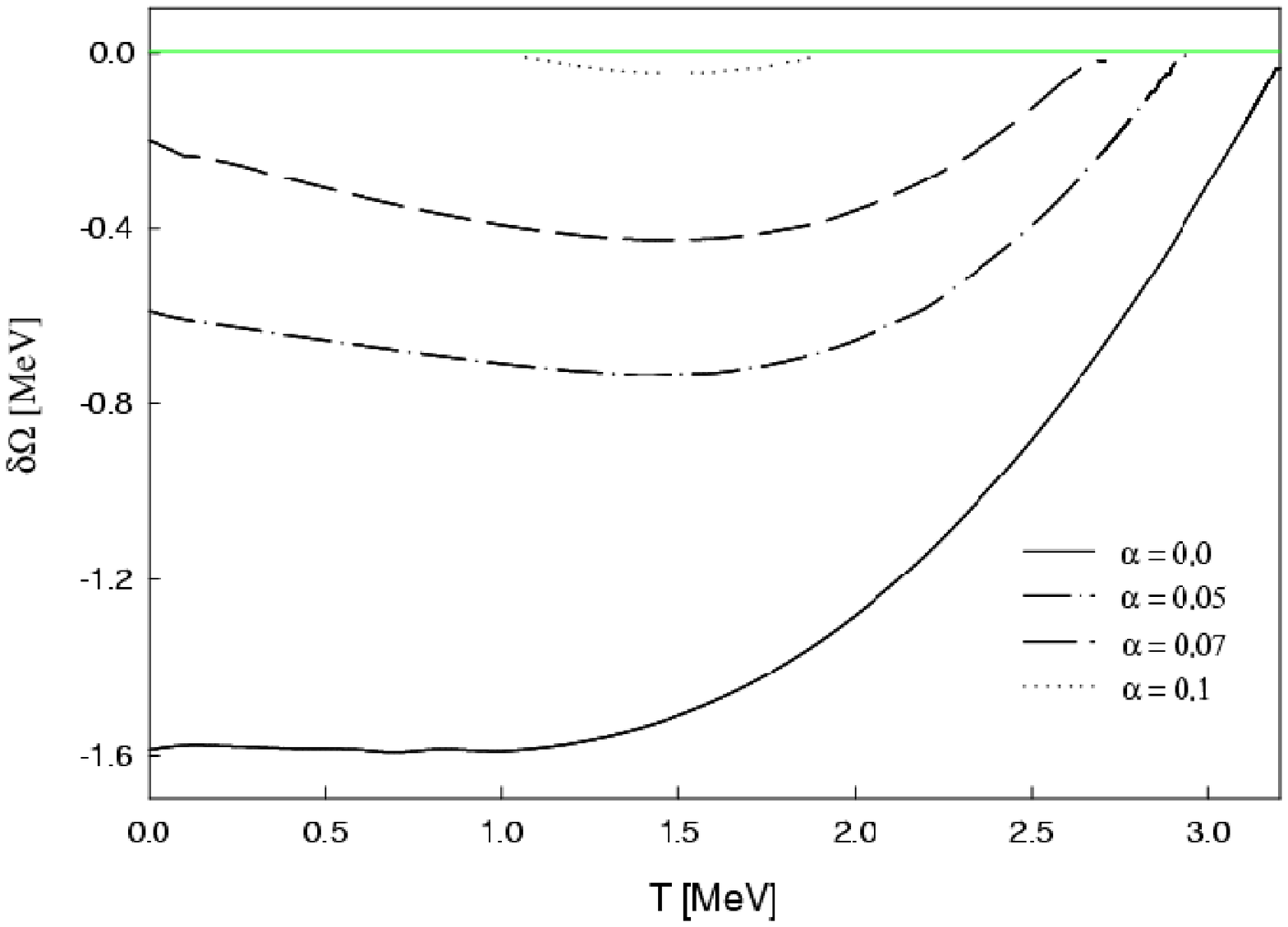,width=2.2in,angle=0}
}
\caption{{\it Left panel.}  Temperature dependence of pairing 
gap for density asymmetries $\alpha = 0.0$ (solid), 0.05 
(dashed-dotted), 0.07 (dashed), and 0.1 (dotted). {\it Right panel.}
Temperature dependence of free energy. Labeling
of asymmetries is as in the left panel.\label{fig:ASC_NO_TRANS}
}
\end{figure}
Minimizing the free energy of an asymmetric superconductor at fixed
density and temperature leads to stable solutions over the entire
region of density asymmetries where non-trivial solutions of
the gap equation exist.\cite{SAL,SL00} 
This can be seen in Fig.~\ref{fig:ASC_NO_TRANS}, where the
temperature and asymmetry dependence of the pairing gap and the
free-energy of a homogeneous asymmetric superconductor are
shown.  In particular, we see that for a fixed temperature, 
the gap and the free energy are single-valued functions of 
the density asymmetry $\alpha$ in the particle-number-conserving 
scheme -- in contrast to what is found in the non-conserving scheme, 
where double-valued solutions appear.

At large asymmetries, the dependence of the gap on the temperature
shows a ``re-entrance'' phenomenon.  As the temperature is increased
from low values at which the asymmetry is too large to sustain
a gap, a critical temperature is reached at which pairing 
correlations take hold.  (For example, this is seen for the 
$\alpha = 0.1$ case in Fig.~\ref{fig:ASC_NO_TRANS}).  This 
behavior can be attributed to the increase of phase-space 
overlap between the quasiparticles that pair, due to the 
thermal smearing of the Fermi surfaces.  Further increase of 
temperature suppresses the pairing gap at a higher critical
temperature due to thermal excitation of the system, in much 
the same way as in the symmetric superconductors.  Clearly, in
this scenario the pairing gap has a maximum at some intermediate 
temperature. The values of the two critical temperatures are 
controlled by different mechanisms. The superconductivity (or 
superfluidity) is destroyed with decreasing temperature at the
lower critical temperature when the smearing of the Fermi 
surfaces becomes insufficient to maintain the required phase-space 
coherence. The upper critical temperature is the analog of the BCS 
critical temperature and corresponds to a transition to (re-entrance
of) the normal state because of thermal excitation.

Another aspect of the asymmetric superconducting state 
is the gapless nature of the excitations.\cite{ALFORD_GAPLESS,RUSTER} 
One may draw an analogy to the non-ideal Bose gas, for which
only a fraction of the particles are in the zero-momentum ground 
state at temperatures below the critical value for Bose
condensation.  The dynamical properties of gapless superconductors, such 
as response to electroweak probes and transport, depend on the 
ratio $\zeta = \Delta/\delta\mu$ in an essential way: for $\zeta > 1$, 
the response of the system is similar to that of an ordinary 
superconductor; in the opposite limit $\zeta < 1$, 
the system's behavior is essentially non-superconducting (see
e.g.\ Ref.~\refcite{JAIKUMAR}).  These features are easily understood 
by examining the excitation spectrum in both limits.

\subsection{Phases with broken space symmetries}
We now turn to a special class of ASC characterized by broken 
global symmetries -- translational, rotational, or both.  In 1964, 
Larkin and Ovchinnikov\cite{LO} and, independently, Fulde and 
Ferrell\cite{FF} (LOFF) discovered that the superconducting 
\index{Larkin, Ovchninikov, Fulde, Ferell (LOFF) phase} 
state can sustain asymmetries beyond the Chandrasekhar-Clogston 
limit if electrons pair with nonzero center-of-mass (hereafter 
CM) momentum. The weak-coupling result for the critical shift 
in the Fermi surfaces for onset of the LOFF phase is 
$\delta\mu_2 = 0.755 \, \Delta(0)\,[>\delta\mu_1 = 0.707\,\Delta(0)]$. 
Since the condensate wave function depends on the CM momentum 
of the pair, its Fourier transform will vary in configuration 
space, giving rise to a lattice structure. The Fulde-Ferrell 
state is predicated on a plane-wave form $\Delta(\vecr) = 
\Delta\, {\rm exp}(-i\vecP\cdot \vecr)$ for the gap function. 
Larkin and Ovchinnikov considered a number of lattice types
and concluded that the body-centered-cubic 
(bbc) lattice is the most stable configuration 
near the critical temperature.  Recent studies of this problem 
in the vicinity of the critical temperature employing 
Ginzburg-Landau theory show that the face-centered-cubic 
(fcc) structure is favored.\cite{BOWERS_RAJAGOPAL}

For illustrative purposes, let us consider the Fulde-Ferrell state, 
in which case the quasiparticle spectrum is given by
\begin{eqnarray}
\label{LOFF_SPEC}
\omega_{\pm}(\vecP,\vecq) = \frac{1}{2m}\left(
\frac{\vecP}{2}\pm\vecq\right)^2 - \mu _{\pm}\,,
\end{eqnarray}
where the upper sign corresponds to neutrons and the lower sign to 
protons. The spectrum (\ref{LOFF_SPEC}) is obtained by 
applying the following transformation to Eq.~(\ref{BRANCHES}): 
$E_S \to E_S + (P^2+p^2)/2m$ and  $E_A \to \delta \mu
\pm \vecp\cdot \vecP$.  Onset of the LOFF phase entails
a positive increase in the quasiparticle kinetic energy
$\propto Q^2$, which disfavors the Fulde-Ferrell state relative
to the BCS state.  However, the anisotropic term
$\propto \vecP\cdot \vecp$, which can be interpreted as a dipole
deformation of the isotropic spectrum, modifies the phase-space
overlap of the fermions and promotes pairing. 
The LOFF phase becomes stable when the increase in the kinetic 
energy required to move the condensate is smaller 
than the reduction in potential energy made possible by the
increase in the phase-space overlap.  The magnitude of the 
total momentum serves as a variational parameter for 
minimization of the ground-state energy of the system.
\begin{figure}[tb] 
\begin{center}
\epsfig{figure=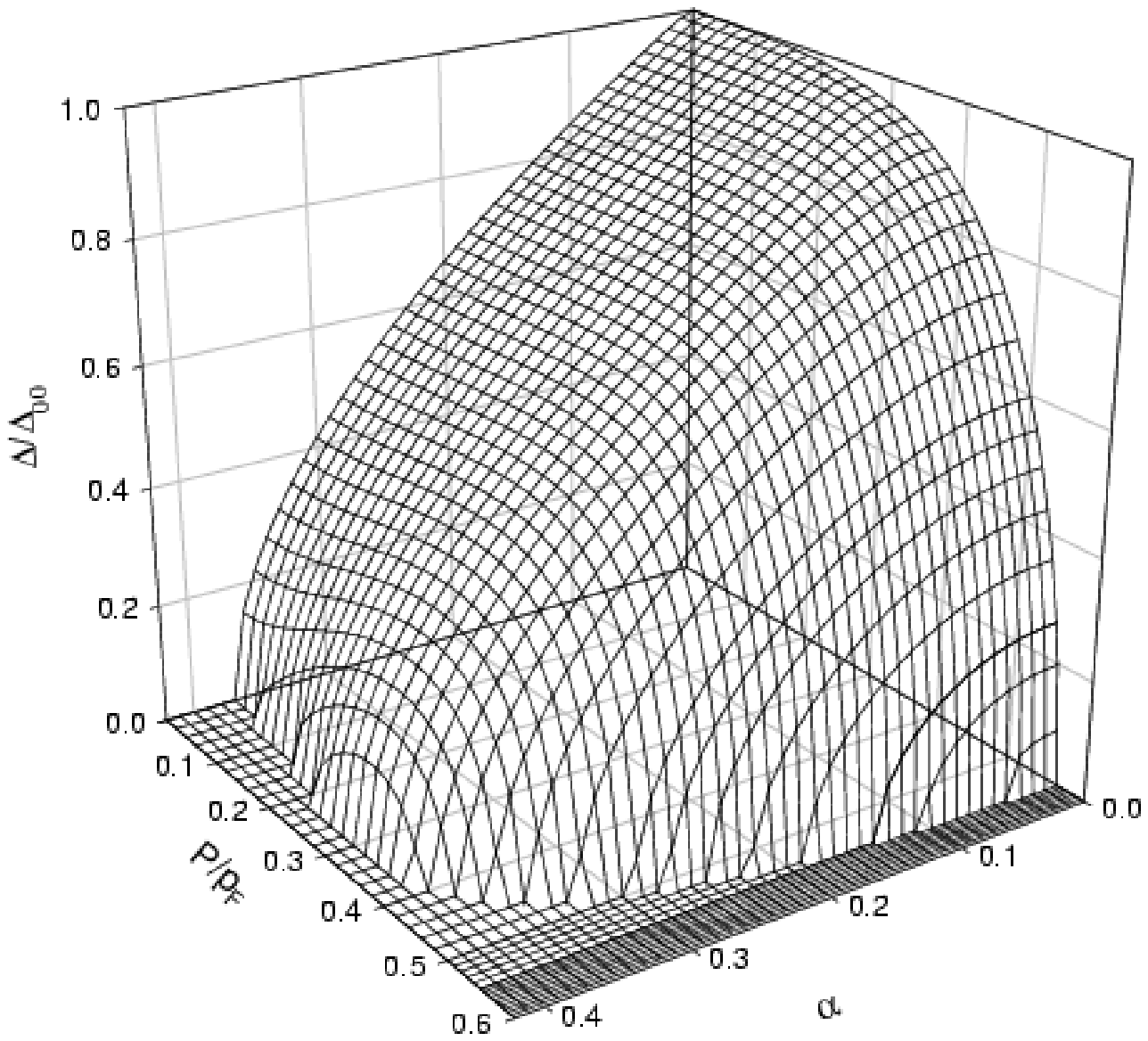,width=5.cm,angle=0}
\hspace{0.7cm}
\epsfig{figure=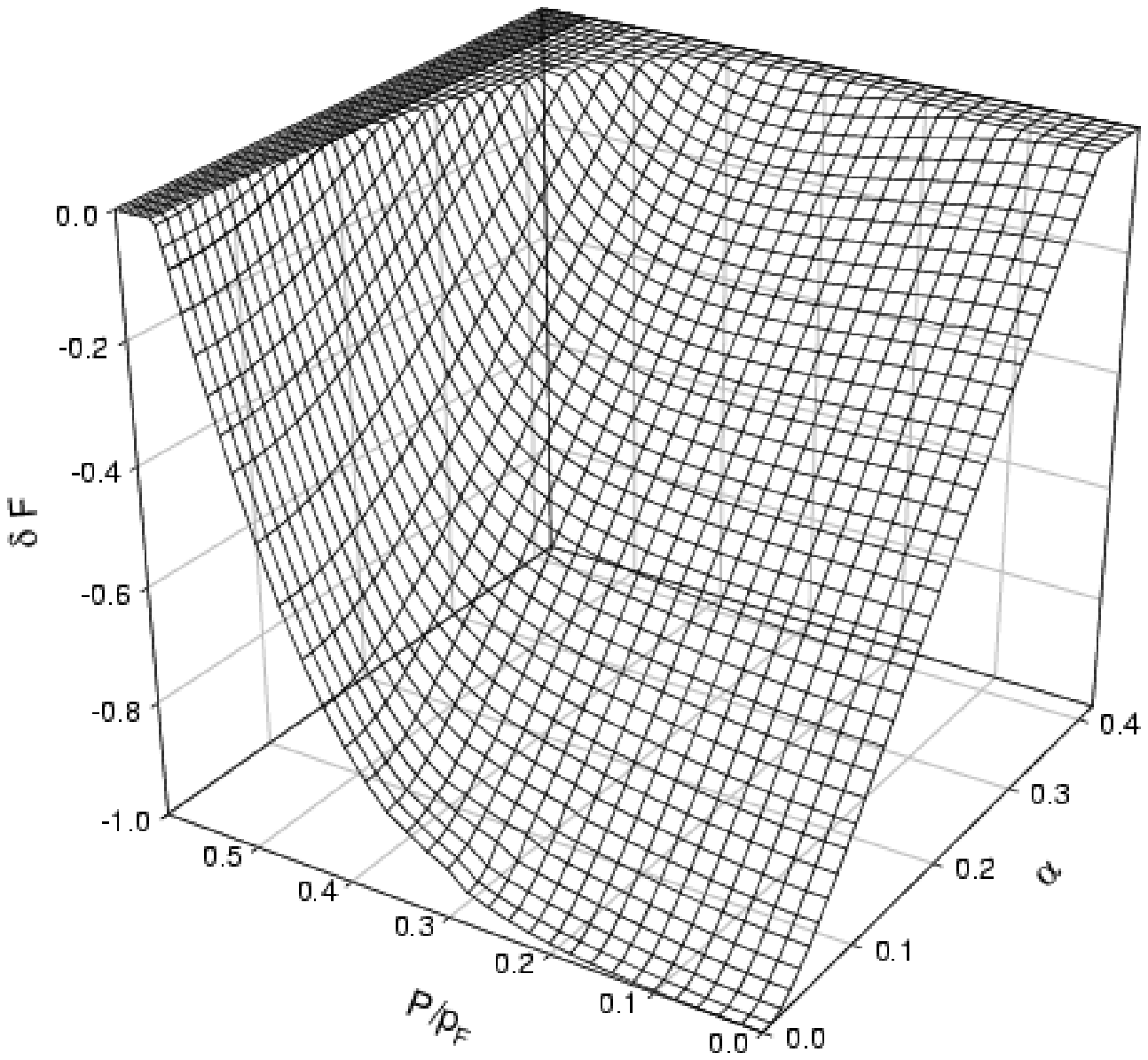,width=5.cm,angle=0}
\caption{{\it Left panel.}
Dependence of pairing gap $\Delta$ in the LOFF phase on density 
asymmetry $\alpha$ and total momentum $P$ of the condensate, relative 
to pairing gap $ \Delta_{00}$ in the limit $P=0=\alpha$.
{\it Right panel.} Dependence of free energy of the 
LOFF phase on the same quantities as in the left panel.}
\label{fig:ASC_LOFF}
\end{center}
\end{figure}
The pairing gap and the free energy of ASC with finite momentum
are shown in Fig.~\ref{fig:ASC_LOFF}. It is assumed that the 
gap function depends parametrically on the {\it magnitude} of the 
CM momentum, but is independent of its direction.\cite{SEDRAKIAN00} 
For such an {\it Ansatz} the anisotropy of the spectrum appears only in the 
Fermi functions in the kernels of Eqs.~(\ref{GAP_PARTIAL}) 
and (\ref{DENS1}) and is averaged through the phase-space 
integration.  It is seen in Fig.~\ref{fig:ASC_LOFF}
that an ASC-LOFF state arises for arbitrary finite momentum 
of the condensate below some critical value.  
For large enough asymmetries the minimum
of the free energy moves from $P=0$ to intermediate values of
$P$, i.e., the ground state of the system corresponds to a 
condensate with nonzero CM momentum of Cooper pairs. 
Note that for the near-critical range of asymmetries, the
condensate exists only in the LOFF state and its dependence on the
total momentum exhibits the re-entrance behavior found in the 
temperature dependence of the homogeneous ASC. The order
of the phase transition from the LOFF to the normal state is a
complex issue that depends on the preferred lattice structure,
among other things (see Ref.~\refcite{Casalbuoni:2003jn} and 
work cited therein).

Due to the Pauli exclusion principle, a noninteracting fermionic 
gas fills an isotropic Fermi sphere; similarly, if there are 
two types of noninteracting fermions, each species fills an
isotropic Fermi sphere. 
Consider now a strongly interacting system that is a Fermi liquid 
rather than a Fermi gas.  According to Fermi-liquid theory, 
the states of the interacting system are reached by switching
the interaction on adiabatically.  Driven by this process,
the noninteracting gas evolves into a strongly-interacting
liquid, in which the dressed single-particle degrees of
freedom -- the quasiparticles -- once again fill a spherical 
shell isotropically.  However, this simple Fermi-liquid picture 
may not hold in two-component (or multi-component) fermionic 
systems in which the fermions of differing species 
interact via strong pairing forces.  Indeed, there can
exist a stable superconducting phase that sustains ellipsoidal
deformations of the Fermi-surfaces, a phase hereafter referred to
as deformed Fermi-surface superconductivity\cite{DFS1,DFS2} (DFS).
\index{Deformed Fermi surface phase}
\index{Fermi surface!deformation}

The quadrupole deformations of the Fermi surfaces are described
by expanding the quasiparticle spectrum in spherical harmonics 
and keeping the $l = 2$ contributions,\cite{DFS1,DFS2}
\begin{eqnarray}
\label{exp}
\omega^D_{\pm}(\vecq)= \omega_{\pm} (\vec q)+ \epsilon_{2,\pm}P_2(x)\,,
\end{eqnarray}
where $\omega_{\pm} (\vec q)$ is the spectrum of the homogeneous 
ASC and the coefficients $\epsilon_2$ describe the deformations of 
the Fermi surfaces that break the rotational O(3) symmetry down to 
O(2).  The O(2) symmetry axis is chosen spontaneously.
Thus, the quasiparticle spectrum of the DFS phase is obtained from 
the spectrum of homogeneous ASC by the transformations 
\be 
E_S \to E_S + (\epsilon_{2,+}+\epsilon_{2,-})/2\mu\,,
\quad E_A \to E_A + (\epsilon_{2,+}-\epsilon_{2,-})/2\mu\,.
\ee
We observe that the leading harmonic term responsible for 
deformation of either Fermi surface is that for $l=2$, not $l=1$, 
since the latter corresponds to a {\it translation} of one 
Fermi sphere relative to the other, without deformation.  The 
deformations are deemed to be stable if they lower the 
free energy of the system relative to its value in the 
undeformed state.

\begin{figure}[tb] 
\begin{center}
\epsfig{figure=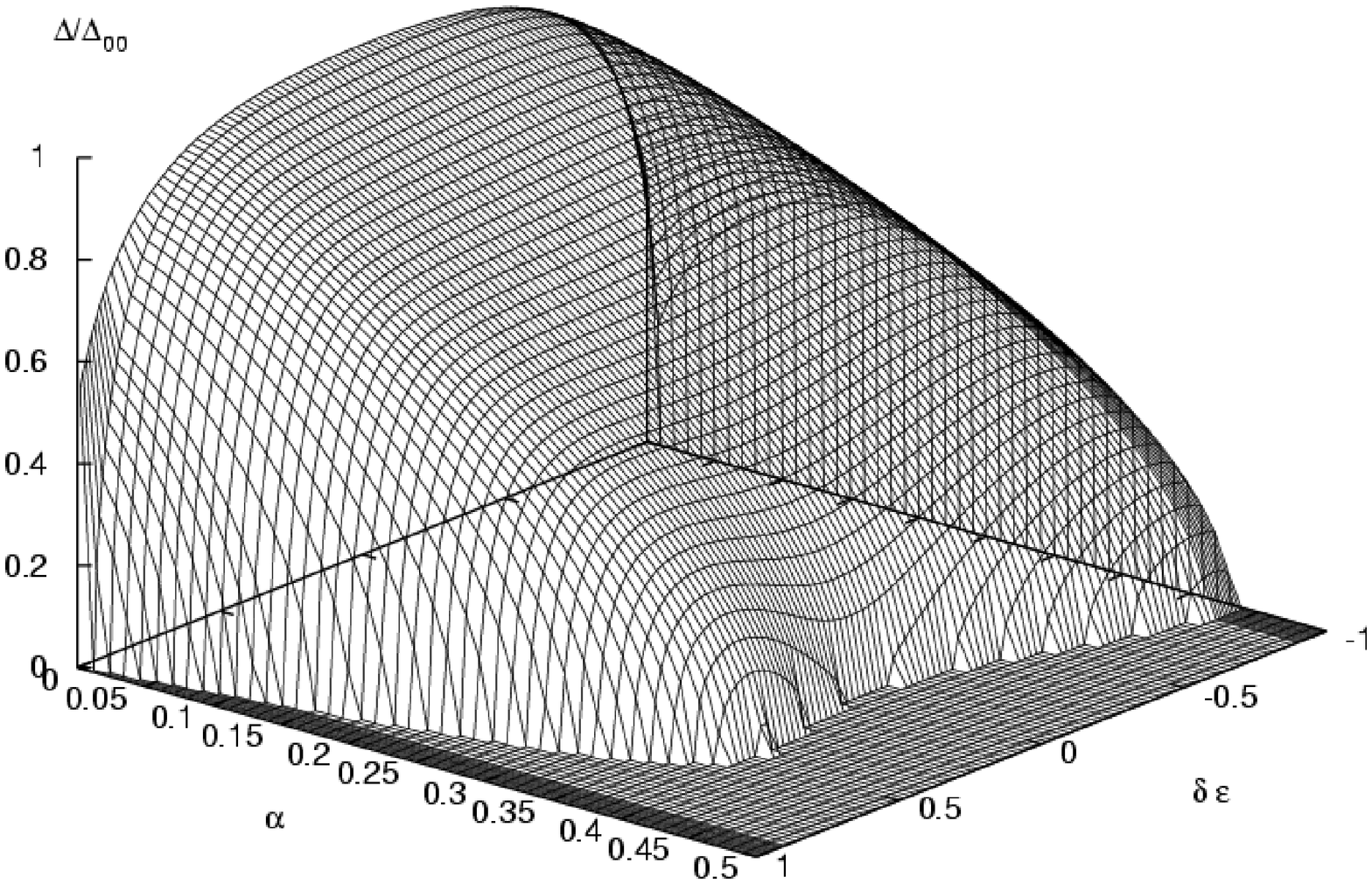,height=2.in,width=2.in,angle=-90}
\hspace{0.7cm}
\epsfig{figure=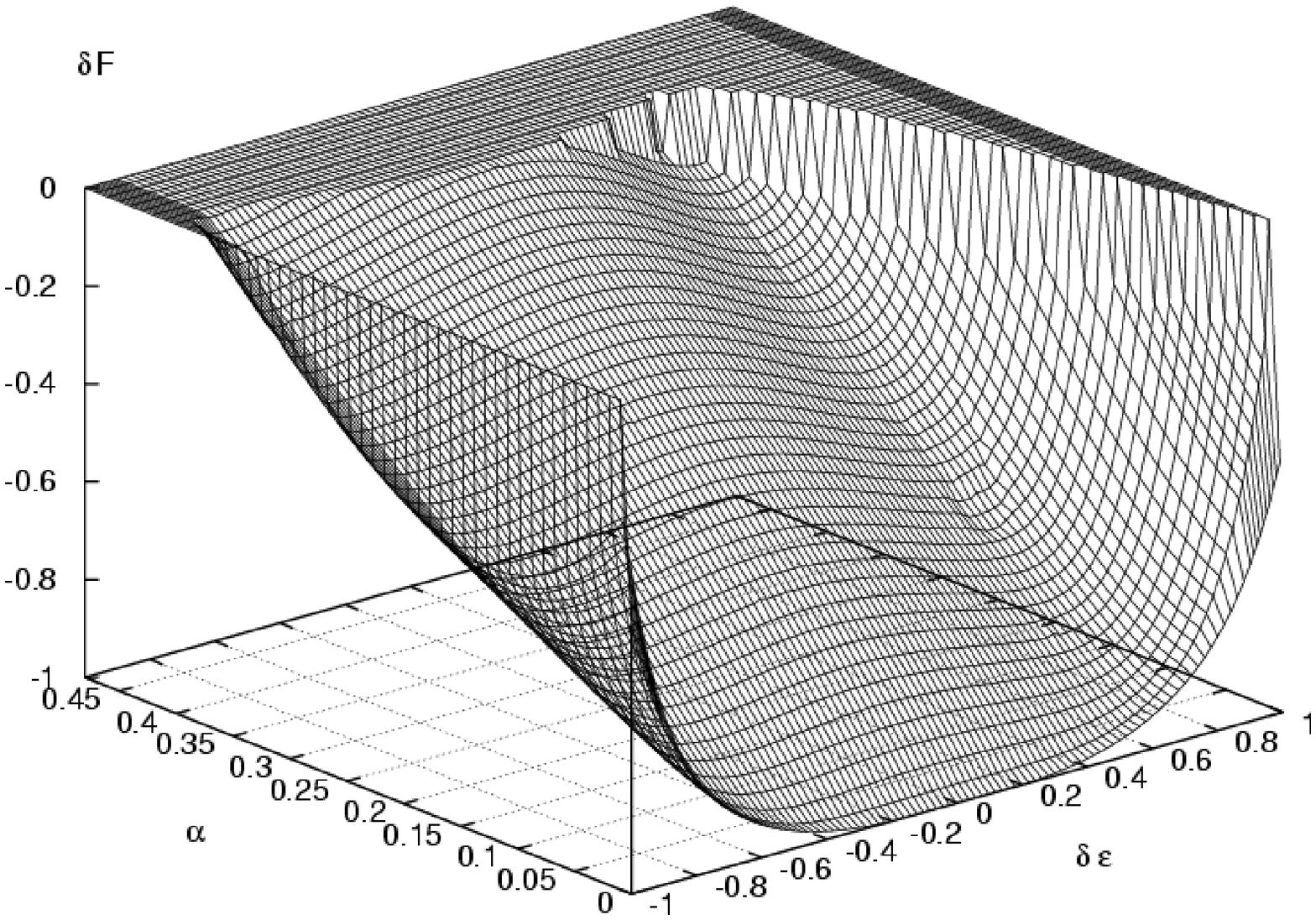,height=2.in,width=2.in,angle=-90}
\caption{
{\it Left panel.}  Dependence of pairing gap in the DFS phase
on density asymmetry and total momentum of the condensate. 
{\it Right panel.}  Dependences of free energy of the DFS phase 
for the same input parameters as in Fig.~5.~
} 
\label{fig:ASC_DFS}
\end{center}
\end{figure}

\begin{figure}[t]
\begin{center}
\includegraphics[height=2.8in,width=2.8in,angle=-90]{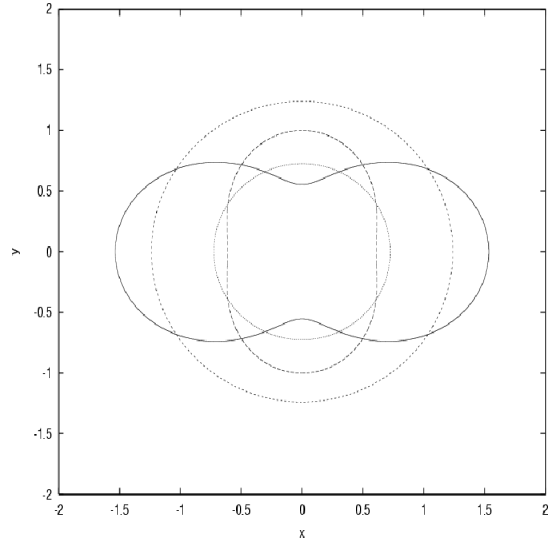}
\caption{
Projection of the Fermi surfaces on a plane parallel to the axis of 
symmetry breaking. The concentric circles correspond to the two 
populations of spin(isospin)-up and spin(isospin)-down fermions 
in a spherically symmetric state ($\delta\epsilon  = 0$), while the 
deformed symbols correspond to the state with relative deformation 
$\delta\epsilon = 0.64$.  The spin(isospin)-density asymmetry is 
$\alpha = 0.35$.}
\label{fig:SHAPES}
\end{center}
\end{figure}
The deformation parameters $\epsilon_{2\pm}$ are determined by
minimization of the free energy of the system, as was done for
the Cooper-pair momentum parameter $P$ in the case of broken
translational invariance (LOFF states).  Three-dimensional plots
of the dependence of the pairing gap and free energy of the 
DFS phase on asymmetry and the relative deformation 
$\delta\epsilon = (\epsilon_{2,+}-\epsilon_{2,-})/2\mu$
are provided in Fig.~\ref{fig:ASC_DFS}.
Fig.~\ref{fig:SHAPES} shows a typical deformed Fermi-surface
configuration that lowers the expected ground-state 
energy below that of the non-deformed state.  At $\alpha = 0$, 
the critical deformation for which pairing ceases is the same 
for prolate and oblate deformations.  At finite $\alpha$ and in
the positive range of $\delta\epsilon$, the maximum value of 
the gap is attained for constant $\delta\epsilon$; at
negative $\delta\epsilon$ the maximum increases as a function 
of deformation and saturates for $\delta\epsilon \simeq 1$. 
The re-entrance phenomenon sets in for large asymmetries as 
$\delta\epsilon$ is increased from zero to finite values. 
(N.B. The essential difference between LOFF and DFS phases is 
that in the latter, the translational symmetry of the 
superconductor remains unbroken.)

To complete our discussion of pairing states in nonrelativistic 
asymmetric superconductors, we
briefly mention some of the alternatives to the LOFF and DFS phases.
One possibility is that the system prefers a phase separation of 
the superconducting and normal phases in real space, such that 
the superconducting phase contains particles with matching
chemical potentials, i.e.\ is symmetric, while the normal phase
remains asymmetric.\cite{CALDAS} 

Equal-spin (-isospin, -flavor) pairing is another option, if the 
interaction between like-spin particles is attractive.\cite{COMB,GUB}  
Since the separation of the Fermi surfaces does not affect 
spin-1 pairing on each Fermi surface, an asymmetric 
superconductor may evolve into a spin-1 superconducting 
state (rather than a non-superconducting state) as the 
asymmetry is increased.  Therefore spin-1 pairing becomes the
limiting case for very large asymmetries.  If the single-particle states 
defining the different Fermi surfaces are characterized by spin 
(as is the case in the metallic superconductors), the pairing 
interaction in a spin-1 state should be $P$-wave and the transition 
is from $S$-wave to $P$-wave pairing.  If the fermions are 
characterized by one or more additional discrete quantum numbers 
(say isospin as well as spin), the transition may occur between 
different $S$-wave phases (e.g. from isospin-singlet to 
isospin-triplet in the case of nuclear matter).  The possibilities
become especially rich in dense quark matter.

\section{Crossover from BCS pairing to Bose-Einstein condensation}
\label{SECTION4}

A crossover from BCS superconductivity to Bose-Einstein
condensation (BEC) is exhibited in fermionic systems with attractive 
interactions under sufficient decrease of the density and/or
sufficient increase of the interaction strength. The transition 
from large overlapping Cooper pairs to tightly bound non-overlapping 
bosons can be described entirely within the ordinary BCS theory, 
if the effects of fluctuations are ignored (mean-field approximation).
Early studies of this type of transition were carried out in the 
contexts of ordinary superconductors,\cite{LEG} 
excitonic superconductivity in semiconductors,\cite{KEL} and, 
at finite temperature, an attractive fermion gas.\cite{NOZ}  
Although the BCS and BEC limits are physically quite different, 
the transition between them is found to be smooth within 
ordinary BCS theory.

Several authors have considered the BCS-to-BEC transition 
in the nuclear context.  In isospin-symmetric nuclear matter,
neutron-proton ($np$) pairing undergoes a smooth transition 
leading from an assembly of $np$ Cooper pairs at higher densities 
to a gas of Bose-condensed deuterons as the nucleon density is 
reduced to extremely low values.\cite{ALM,SCK,SSALR,LNSSS,SC}  
This transition may be relevant -- and could then yield valuable 
information on $np$ correlations -- in low-density nuclear systems
(especially the nuclear surface), in expanding nuclear matter 
from heavy-ion collisions, and in supernova matter.  The 
underlying equations of the theory are (\ref{GAP_PARTIAL}) 
and (\ref{DENS1}) with $E_A = 0$; we shall address 
the effects of asymmetry at a later point.
\begin{figure}[t]
\begin{center}
\epsfig{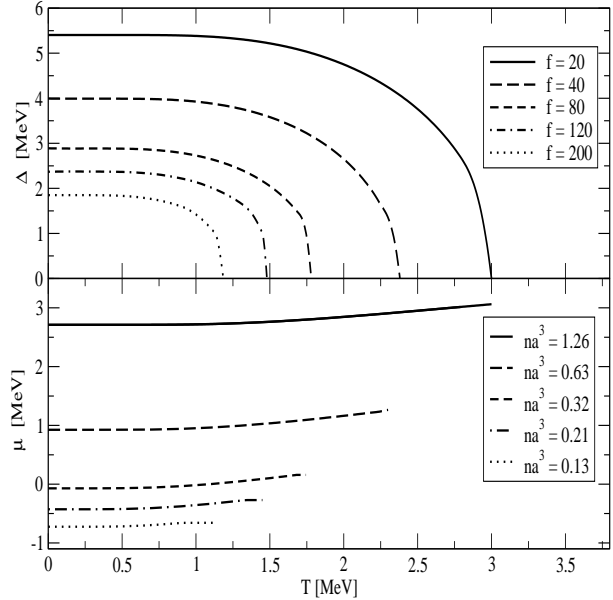}
\caption{
Dependence of pairing gap ({\it upper panel}) and chemical
potential ({\it lower panel}) on temperature for fixed values 
of the ratio $f = n_0/n$, where $n$ is the baryon density 
and $n_0 = 0.16$ fm$^{-3}$ is the saturation density of 
symmetrical nuclear matter.  Values of the diluteness parameter 
$na^3$ assume a scattering length $a = 5.4$ fm.}
\label{fig:GAP_MU}
\end{center}
\end{figure}
The top panel of Fig.~\ref{fig:GAP_MU}
shows the dependence of the gap function on temperature for 
several densities $n$, given in terms of the ratio $f = n_0/n$,
where $n_0 = 0.16$ fm$^{-3}$ is the saturation density of 
symmetrical nuclear matter.  The bottom panel shows the 
associated chemical potentials $\mu$ computed self-consistently 
from Eq.~(\ref{DENS1}). The low- and high-temperature
asymptotics of the gap function are well described by the BCS
relations $\Delta(T\to 0) = \Delta(0) -
[2\pi\alpha\Delta(0)T]^{1/2}\, {\rm exp}(-\Delta(0)/T)$
and $\Delta(T\to T_{c2})  = 3.06\,\beta [T_{c2} (T_{c2} -T)]^{1/2}$,
respectively, where $T_{c2}$ is the critical temperature of the phase
transition. However, the BCS weak-coupling values $\alpha = 1 = \beta$
must be replaced with $\alpha \sim 0.2$ and $\beta \sim 0.9$.  
As a consequence, the ratio of the zero-temperature 
gap to the critical temperature deviates from the familiar
BCS result $\Delta(0)/T_{c2} = 1.76$.
Deviations from the original BCS theory are understandable in that
(i)~the system is in the strong-coupling regime, and
(ii)~the pairing is in a spin-triplet rather than a spin-singlet
state.~\cite{SC}

One measure of coupling strength is the ratio 
$ \Delta(0)/\vert \mu\vert $ of the zero-temperature energy gap 
to the magnitude of the chemical potential.
It is seen from Fig.~\ref{fig:GAP_MU} 
that the strong-coupling regime is realized 
for $f\ge  40$ (i.e., $\Delta \gg \mu$).  At $f =20$ the 
system is in a transitional regime ($\Delta \sim \mu$).
Another measure of coupling is the diluteness parameter $n|a|^3$, 
where $a$ is the scattering length. In agreement with the first
criterion, the matter is in the dilute (or strong-coupling) regime 
for $f\ge  40$, since this range corresponds to $na^3 = 0.63  < 1$
when $a$ is taken as the triplet $n-p$ scattering length,
5.4 fm.  A signature of the crossover from weak to strong coupling 
is the change of sign of the chemical potential, which
occurs for $f \approx 80$ (Fig.~\ref{fig:GAP_MU}), slightly below the
crossover density between weak-coupling and strong-coupling regimes.

In the limit of vanishing
density, $f\to \infty$, the value of the chemical potential at $T=0$
tends to $\mu (\infty) = -1.1$ MeV, which is half the binding energy of
the deuteron in free space. Indeed, in this limit the gap equation
reduces to the Schr\"odinger equation for a two-body bound state,
with the chemical potential assuming the role of the energy
eigenvalue.\cite{LEG}  Thus, the BCS condensate of Cooper pairs in the
$^3S_1$--$^3D_1$ state evolves into a Bose-Einstein condensate of 
deuterons as the system crosses over from the weak- to the strong-coupling 
regime.  The crossover is smooth, taking place without change of symmetry
of the many-body wave function.

How does isospin asymmetry affect the transition?  As the system 
is diluted, the critical asymmetry at which the pairing disappears 
increases from small values of the order of 0.1 up to the asymptotic 
value $\alpha = 1$.  The reason for this behavior is that in the 
low-density matter, the excess neutrons do not appreciably change 
the wave functions of protons, which are bound into pairs.\cite{LNSSS}  
At asymptotically small densities,
the chemical potential of protons tends to $\mu_p (\infty) = -2.2$ MeV,
which is just the binding energy of the system per half the number 
of particles bound into deuterons.  The chemical potential 
of neutrons is determined by the excess particles in the continuum 
and goes asymptotically as $\mu_n (\infty) \to 0$ (i.e.\ there is 
ultimately no energy cost in adding a neutron to the system).  Note 
that the asymptotic behavior described is independent of the 
degree of isospin asymmetry.

In closing this section, we call attention to the remarkable
progress achieved during the last few years in trapping and 
manipulating ultracold fermion gases.  The strength of the 
two-body interaction between the constituent fermionic atoms can 
be tuned using the Feshbach resonance mechanism, 
\index{Feshbach resonance} by varying 
the external magnetic field\cite{FESHBACH1,FESHBACH2,FESHBACH3};
thus, the entire range from weak to strong couplings can be probed. 
Recent experiments on ultracold atomic gases have begun to
explore their properties in cases where pairing occurs between
atoms in different hyperfine states, which are unequally
populated.\cite{ZWI,HULET}  Systems of this kind are also
subject to intensive theoretical study, with specific
attention to homogeneous ASC phases,\cite{ATOMS_THEORY1} 
phases with broken symmetries\cite{ATOMS_THEORY2} and their 
realization in finite trap geometries~\cite{ATOMS_THEORY3}.
The universal features of ASC revealed by this effort should
contribute significantly to our understanding of nucleonic
pairing under isospin-asymmetric conditions.

\section{Vortex states in compact stars}
\label{SECTION5}
\subsection{Currents and quantized circulation}

\index{vortex states}
The macroscopic physics of neutron-star rotation and its anomalies
observed in the timing of pulsars can be described within the 
hydrodynamic theory of superfluids suitably extended to multifluid 
systems~\cite{VS,MENDELL_LINDBLOM,MENDELL,SS95,MENDELL2,PRIX,ANDERSSON}.  
The elementary constituents of 
a Fermi superfluid -- the Cooper pairs -- are characterized by a
coherence length $\xi$.  
\index{coherence length}
On length scales $L\gg \xi$, the condensate of 
Cooper pairs can be described by a single wave function $\psi$, and 
the condensate forms a macroscopically coherent state. 
At an intermediate or ``mesoscopic'' scale, stellar rotation and the
presence of a magnetic field lead to the formation of vortices, 
\index{rotational vortices}
macroscopic quantum objects whose distinctive property is the 
quantization of circulation around a path encircling the vortex
core.  Since the condensate wave function must be single-valued
at each point of the condensate, the circulation is quantized 
in units of $2\pi\hbar$.
On writing $\psi = \psi_0 e^{i\chi}$,
the gauge-invariant superfluid velocities can be expressed through
the gradient of the phase of the superfluid 
order parameter $\chi$ and the value of the vector potential ${\bm A}$:
\begin{equation}\label{eq:ASv}
{\bm v}_{\tau}=\frac{\hbar}{2m_{\tau}}{\bm\nabla }
\chi _{\tau}-\frac{e_{\tau}}{m_{\tau}c} {\bm A}\,.
\end{equation}
In this expression, $ e_{\tau} \equiv (e, 0 )$ specifies the 
electric charge of protons ($p$) and neutrons ($n$) respectively, 
$m_{\tau}$ is their bare mass, and $\tau$ stands for $n$ or $p$.
Applying the curl operator to Eq.~(\ref{eq:ASv}) and implementing 
quantization of the circulation (with the phase of the superfluid
order parameter changing by $2\pi$ around a closed path), one finds
\begin{eqnarray}\label{eq:AS:curlv}
\mathop{\rm curl} {\bm v_{\tau}} = \frac{\pi \hbar}{m_{\tau}}
{\bm \nu}_{\tau} \sum_{j} \delta^{(2)} ({{\bm x}}-{{\bm x}}_{\tau j})
- \frac{ e_{\tau}}{m_{\tau}c}{\bm B} \equiv {\bm \omega}_{\tau}\,,
\end{eqnarray}
where $\pi \hbar/m_{\tau}$ is the quantum of circulation, ${\bm \nu}_{\tau}
\equiv {\bm \omega}_{\tau}/\omega_{\tau}$ is a unit vector
along a given vortex line, ${{\bm x}}_{\tau j}$ defines the position
of a vortex line in the plane orthogonal to the vector
${\bm \nu}_{\tau}$, $\delta^{(2)}$ is a two-dimensional Dirac delta
function in this plane, and ${\bm B}=\mathop{\rm curl} \, {\bm A}$ 
is the magnetic-field induction.
The index $j$ is summed over the sites of vortex lines.  
Eq.~(\ref{eq:AS:curlv}) treats the vortex cores as singularities 
in the plane orthogonal to $ \vecnu_{\tau}$; this simplification is 
justified on scales larger than the coherence length of the condensate.
For a single vortex, the integral of Eq.~(\ref{eq:AS:curlv}) completely 
determines the superfluid pattern.  Since this equation is linear,
the superfluid pattern created by a larger number of vortices 
is formed by superposition of the flows induced by each vortex.  
Obviously, the resulting net flow depends on the arrangement 
of the vortices. 

The condensate wave function can be written as $\psi({\bm x})=
f(r) e^{i\theta}$ in cylindrical polar coordinates $(r,\theta, z)$. 
Upon integrating Eq.~(\ref{eq:AS:curlv}), the neutron and proton 
superfluid velocities then become
\begin{eqnarray}\label{eq:AS:vn}
{\bm v}_n(r) = \frac{\hbar}{2m_nr}\hat \theta\,, \quad\quad 
{\bm v}_p(r) = \frac{\hbar}{2m_p\lambda}
K_1\left(\frac{r}{\lambda}\right)\hat \theta\,,
\end{eqnarray} 
where $K_1$ is the Bessel function of imaginary argument.
The divergence of the neutron-vortex velocity ${\bm v}_n(r)$ 
as $r\to 0$ is regularized by a cutoff $\Lambda\sim \xi_n$.  
The long-range nature of ${\bm v}_n(r)$ results in  
slow falloff of a density perturbation in the condensate.
In a proton vortex, the supercurrent is screened exponentially on 
length scales of the order of the penetration depth $\lambda$.
Thus, for $r\gg \lambda$, $K_1(r/\lambda) \simeq {\rm exp} (-r/\lambda)$.

On global, hydrodynamic scales, transition to a continuum 
vortex distribution can be carried through on the right-hand side of  
Eq.~(\ref{eq:AS:curlv}) by defining vortex densities $n_{\tau}=
\sum_{j} \delta^{(2)} ({\bm x}-{\bm x}_{\tau j})$.  Since the 
curl of ${\bm v}_n$ is simply $2\Omega$ for rigid-body rotations, 
the number density of vortices in the neutron superfluid is 
related to the macroscopic angular velocity of the neutron 
condensate by the Feynman formula 
\begin{eqnarray}\label{eq:AS:nn}
n_n = \frac{2m_n\Omega}{\pi\hbar}\,.
\end{eqnarray}
For typical pulsar periods $P$ in the range $0.05 < P < 0.5$ s, 
one has $n_n\simeq 6.3.\times 10^3 ~ P^{-1}\sim 10^4$--$10^5$ per cm$^2$.
In the case of a charged superfluid, Eq.~(\ref{eq:AS:curlv}) can 
be transformed to a contour integral over a path along which ${\bm v}_p=0$, 
since the supercurrent is screened beyond the magnetic field 
penetration depth $\lambda$.  \index{penetration depth}
If the proton superfluid is a
type-II superconductor (i.e., $\lambda/\xi_p > 1/\sqrt{2}$), 
\index{superconductor!type-II}
the continuum vortex limit leads to the estimate
\begin{eqnarray}\label{eq:AS:np}
n_p =\frac{B}{\Phi_0}\simeq  5\times 10^{18}
~{\rm cm}^{-2}\,,
\end{eqnarray}
where $\Phi_0 = \pi\hbar c/e$ is the flux quantum.  We
note that the number of proton vortices per neutron vortex is
$n_p/n_n\sim 10^{13}-10^{14}$, independently of their arrangement.
The energy of a bundle of neutron or proton vortices is 
minimized by a triangular lattice with a unit cell area 
$n_{\tau}^{-1}=(\sqrt{3}/2)\,d_{\tau}^2$.
The lengths of ``basis vectors'' of the lattices in the neutron 
and proton condensates (the inter-vortex distances) are
\begin{eqnarray}
   d_n= \left(\frac{\pi\hbar}{\sqrt{3}\, m_n\, \Omega}   \right)^{1/2}\,,
\quad \quad  d_p=\left(\frac{2\, \Phi_0}{\sqrt{3} \, B}   \right)^{1/2}\,,
\end{eqnarray}
where $B$ is the mean magnetic-field induction.
Using the estimates given in Eqs.~(\ref{eq:AS:nn}) and (\ref{eq:AS:np}), 
one finds that the neutron and proton inter-vortex distances are
$d_n \sim 10^{-2}-10^{-3}$ cm and $d_p\sim 10^{-9}$ cm, respectively.
For typical values of the microscopic parameters, the penetration 
depth is of the order $100~{\rm fm} = 10^{-11}$ cm.  Therefore 
the conditions $\xi_n\ll d_n$ and $\xi_p\ll {\rm min }(\lambda,\,d_p)$ 
are satisfied, and the use of hydrodynamics on the local scale 
is valid.  It is also clear that global hydrodynamics can be 
applied on scales that are much larger than $d_n$ (a fraction 
of millimeter).  

The strong interaction between the neutron and proton fluids gives
rise to the {\sl entrainment effect}: the supercurrent is a linear 
combination of the velocities of both fluids.\cite{VS,ALPAR1} 
More specifically, the mass current and velocity vectors are 
related by a nondiagonal density matrix in isospin space,
\be \label{0}
\left(\begin{array}{c} {\vecp}_1 \\
{\vecp}_2 \end{array}\right)  
=\left(\begin{array}{cc} \rho_{11}& \rho_{12}\\ 
\rho_{21}& \rho_{22}\end{array}\right) 
\left(\begin{array}{c} {\vecv}_ 1 \\ {\vecv}_ 2 \end{array}\right)\,, 
\ee 
where 1 and 2 label the isospin projections.  
The off-diagonal elements, which would vanish in the noninteracting 
limit, are evidently responsible for the entrainment effect.
One fundamental consequence of this effect is that the neutron vortex 
carries a non-quantized magnetic flux\cite{VS,ALPAR1} of the same 
order of magnitude as the flux quantum $\Phi_0$.  If the proton fluid 
forms a type-II superconductor, the number of proton vortices (sometimes
called flux tubes) is $10^{12}-10^{13}$ {\it per} neutron vortex 
[see Eq.~(\ref{eq:AS:np})].  Accordingly, the neutron-vortex 
motion (dynamics) is likely to be affected by the proton-vortex array. 

The arrangement of the proton-vortex lattice is a complex issue, and
there exist several models for its configuration.
\begin{itemize}
\item[(i)]
A class of ({\sl flux-tube}) models envisions the proton-vortex array 
to be spatially homogeneous, the vorticity vector being inclined 
by some angle with respect to the spin vector.\cite{RUDERMAN}
Further, it is assumed that the flux tubes  act as extended
pinning centers for neutron vortices. In such 
models, a change in the neutron-vortex distribution is achieved
by vortex creep of neutron vortices through the array of flux tubes.

\item[(iia)]
{\sl Vortex cluster} models predict clustering of proton vortices 
over about $10\%$ of the area occupied by a neutron vortex.  
One particular model generates a bundle of proton vortices coaxial 
with the neutron vortex, through the entrainment currents induced by 
neutron-vortex circulation.\cite{SS1}  This gives rise to 
an average axisymmetrical magnetic field whose magnitude is compatible 
with pulsar observations. \index{vortex clusters}

\item[(iib)] The homogeneous distribution of proton vortices could 
be generically unstable towards phase separation between a phase 
containing dense mesh of proton vortices and a phase devoid of vortices.
A necessary condition is that the vortex lattice is sufficiently dilute, 
the mean intervortex distance being much larger than the penetration 
depth.\cite{MUZIKAR}

\item[(iic)]
Proto-neutron stars are likely to possess natal magnetic fields; 
the nucleation of such a field will be associated with a first-order 
normal-superconducting phase transition, squeezing the field 
into bubbles of superconducting regions with high $B\sim 10^{14}$ G, 
and forming stable protonic vortex arrays 
which again cover about $10\% $ of the total area.\cite{SED_CORDES}  
The dynamics of the neutron-vortex array in vortex-cluster models 
is controlled by the electromagnetic scattering of electrons 
off a vortex cluster.
\end{itemize}

Current models of BCS pairing of protons do not exclude the 
possibility that there is a transition from type-II to type-I 
superconductivity of protons as the density is increased.\cite{SSZ}
Type-I superconducting protons will have domain structures analogous 
to those observed in laboratory experiments on terrestrial
superconducting materials.  The electrodynamics of the proton 
domain structures in NS can be treated by adapting the theories 
developed for laboratory superconductors, in which the magnetic 
fields are generated by normal currents driven around a cylindrical 
cavity by temperature gradients.\cite{GZ}  A recent theoretical 
study\cite{BUCKLEY} examined the effect of interactions between 
neutron and proton Cooper pairs on the status of proton 
superconductivity.  The results suggest that type-I 
superconductivity can be enforced throughout the entire stellar 
core if the strength of the interaction between Cooper pairs 
is significant.  However, within the mean-field BCS theory, 
the Cooper pairs are noninteracting entities, and any deviations 
from this picture must be due to fluctuations.  Alford 
et~al.\cite{ALFORD} estimated the strength of the interaction 
between neutron and proton Cooper pairs due to fluctuations 
and found it to be too small to account for the interactions 
assumed in Ref.~\refcite{BUCKLEY}.  Nevertheless, type-I 
superconductivity of protons is not excluded by current 
calculations of $^1S_0$ pairing of protons, and we 
shall address below its potential implications for the 
macroscopic manifestations of superfluidity in neutron stars.

\subsection{Constraints placed by neutron-star precession 
on the mutual friction between superfluid and normal-fluid
components}

Since most of the inertia of a NS is carried by the neutron superfluids 
in the core and in the crust, the key to an understanding of NS 
rotational anomalies lies in the transfer of angular momentum
from the superfluid to the normal (unpaired) component of the star, 
whose rotation is observed through the magnetospheric emission.  At 
the local hydrodynamical scale, the rate of angular momentum transfer 
between the superfluid and normal components is determined by 
the equation of motion of a superfluid neutron vortex line.  In 
the approximation that the inertial mass of the vortex is 
neglected, the equation of motion is 
\be\label{FORCE_BALANCE} 
\omega_S (\vecv_S-\vecv_L)\times\vecnu + 
\zeta (\vecv_L-\vecv_N) +\zeta' 
(\vecv_L-\vecv_N)\times \vecnu  =0\,,
\ee
where $\vecv_S$ and $\vecv_N$ are the superfluid and normal 
fluid velocities, $\vecv_L$ is the velocity of the vortex,
$\vecnu$ is a unit vector along the vortex line, $\omega_S$ 
is the unit of circulation, and $\zeta$, $\zeta'$ are 
(dimensionless) friction coefficients, also known as the 
drag-to-lift ratios.  These coefficients encode the 
essential information on the microscopic processes of 
interaction of vortices with the ambient unpaired fluid. 
Microscopic calculations commonly indicate $\zeta' \approx 0$, 
and one is left with a single parameter $\zeta$.

In the NS crust, neutron vortices are embedded in a lattice 
of neutron-rich nuclei, and the $\zeta$ coefficient is determined
by the interaction of the vortices with the nuclei and the electron 
plasma.  (In some models, neutron vortices are localized -- pinned 
to the nuclei or situated in between them.\cite{ALPAR2,LINK}
If the pinning is strong, Eq.~(\ref{FORCE_BALANCE}) is not 
valid, since the forces acting on the vortex are not linear 
functions of velocities.  However, the regime of perfect
pinning can be identified with the $\zeta\to \infty$ limit.)
In the core of the star, the friction is controlled by the 
interaction of neutron vortices with the ambient electron-proton 
plasma; Eq.~(\ref{FORCE_BALANCE}) is valid under these conditions. 
Initial studies of the dynamical coupling between the superfluid 
and the normal fluid focused on interpretation of the observed
post-glitch relaxation of pulsar rotational periods.  It turns 
out that such interpretation is fraught with ambiguity, because 
the long relaxation times can be obtained in the two 
opposite limits of weak ($\zeta\to 0$) and strong 
($\zeta\to\infty$) couplings.\index{pulsars!post-glich relaxation}

Recent observation\cite{PRECESSION1} of long-term periodic 
variabilities in PSR B1828-11, if attributed to precession 
of this pulsar, challenges existing theories of vortex dynamics 
in NS.\cite{SHAHAM,SWC,BLINK}  \index{pulsars!precession}
The importance of the inferred
precession mode stems from the fact that it involves 
non-axisymmetric perturbations of the rotational state,
removing the degeneracy with respect to $\zeta$ that is inherent 
in the interpretation of post-glitch dynamics.  In the frictionless 
limit, a star must precess at the classical frequency $\epsilon\Omega$, 
where $\epsilon$ is the eccentricity and $\Omega$ is the rotation 
frequency.  Clearly, then, there must exist a crossover from 
damped to free precession as $\zeta$ is decreased.  The crossover 
is determined by the dimensionless parameters ($I_S/I_N)\beta$ 
and ($I_S/I_N)\beta'$.  Here, $\beta = \zeta/[(1-\zeta')^2+\zeta^2]$, 
$\beta' = 1- \beta(1-\zeta')/\zeta$, $I_S$ is the moment of inertia 
of the superfluid, and $I_N$ is the moment of inertia of the 
crust plus any component coupled to it on time scales much 
shorter than the precession time scale.  The precession 
frequency is\cite{SWC} 
\be\label{PRECESSION}
\Omega_P = \epsilon\Omega_S\left[ \left(1+\beta'\frac{I_S}{I_N}\right)
+i \beta \frac{I_S}{I_N}\right]\,,
\ee
where $\Omega_S$ is the spin frequency and $\epsilon$ is the eccentricity.
A no-go theorem\cite{SWC} states that Eulerian precession in a superfluid 
\index{pulsars!precession!no-go theorems}
neutron star is impossible if ($I_S/I_N)\zeta >1$ (assuming as before 
$\zeta'\to 0$).  There is a subtlety in this result:  the precession 
is impossible because the precession mode, apart from being damped, 
is renormalized by the non-dissipative component of the 
superfluid/normal-fluid interaction ($\propto \beta')$.  In 
effect, the value of the precession eigenfrequency drops below the 
damping frequency for any $\zeta$ larger than the crossover value.  
This counterintuitive result cannot be obtained from arguments based 
solely on dissipation. In fact, according to Eq.~(\ref{PRECESSION}), 
the damping time scale for precession increases linearly with $\zeta$, 
and in the limit $\zeta\to \infty$ one would (wrongly) predict undamped 
precession.  If a neutron star contains multiple layers of superfluids, 
the picture is more complex, but the generic features of the crossover 
are the same.\cite{SWC}  While it is common to study perturbations 
from the state of uniform rotation, the precessional state may 
actually correspond to the local energy minimum of an inclined 
rotor if there is a large enough magnetic stress on the
star's core.\cite{WASSERMAN,AKGUN} 

In the core of a neutron star, the quantized neutron-vortex array is 
embedded in an electron-proton plasma, with the protons in a superconducting
state.  Electrons will scatter off the anomalous magnetic moments of 
(ungapped) neutron quasiparticles localized in the core of a neutron
vortex.\cite{FEIBELMAN}  Because of the $^3P_2$ spin-1 nature of
the order parameter of the neutron superfluid in the core, the 
$^3P_2$ vortex core has an additional magnetization that scatters 
electrons more effectively.\cite{SAULS}  An even more efficient
scattering mechanism comes into play due to the flux $\sim \Phi_0$ 
induced by the proton supercurrent on the neutron vortex
via the entrainment effect.\cite{ALPAR1}  
If the protons are non-superconducting
in some regions of the core, then the strong nuclear interaction
between protons and neutron quasiparticles localized within a vortex
core leads to an efficient coupling of the electron-proton plasma
and the neutron superfluid.\cite{SEDRAKIAN98}  The above models belong to 
the class of weak-coupling theories, i.e.\ $\zeta \ll 1$, and, according 
to the no-go theorems, are compatible with free precession of the neutron 
star. However, these theories of mutual friction assume (unrealistically) 
that the proton-vortex lattice has no effect whatsoever on neutron-vortex 
dynamics in the core.

The mechanism underlying mutual friction in the flux-tube models is 
the slow motion of neutron vortices through the pinning barriers 
(here, flux tubes) via thermally activated creep.  Since, in 
general, the creep models presume that the vortex lattice 
closely follows the rotation of the pinning centers (N.B.\ 
the case of perfect pinning corresponds to $\zeta\to \infty$), 
the effective friction in these models is large,\cite{CHENG}
$\zeta \gg 1$.  The kelvon-vortex coupling in the core provides
another interaction channel, leading again to\cite{BLINK} $\zeta \gg 1$.
Similarly, for vortex-cluster models, in which the neutron-vortex 
lattice and the associated proton-vortex cluster move coherently, 
electron scattering by proton-vortex clusters also gives\cite{SS95}
$\zeta \gg 1$.  Accordingly, these theories are incompatible with 
free precession of a neutron star.  This conclusion has been 
stressed by Link,\cite{BLINK} and it has been argued
that type-I superconductivity could be an alternative.  

In the crust of a neutron star, the neutron-vortex lattice is embedded
in a lattice of nuclei and the charge-neutralizing background created
by an almost homogeneous electron sea.  In the vortex-creep models, 
the neutron-vortex lattice maintains rotational equilibrium 
via thermal and quantal creep through the pinning barriers 
(nuclei).\cite{ALPAR2,LINK}  Hence these models imply $\zeta \gg 1$ and are 
incompatible with free precession.  If the pinning is absent,
either because re-pinning cannot be achieved in post-jump 
equilibrium~\cite{REPINNING}
and/or because of mutual cancellation of the forces from different 
pinning centers, the freely flowing neutron-vortex lattice interacts with
the electron-phonon component of the crust. These interactions are weak 
and lead to\cite{BAYM_EPSTEIN,JONES} $\zeta \ll 1$.  Note that 
the above estimates assume that the ratio $I_S/I_N$ is roughly of 
order unity.  While it is difficult to estimate this ratio
precisely, it is unlikely to differ from unity by many 
orders of magnitude.

\subsection{Type-I superconductivity in neutron stars}

\index{superconductor!type-I}
The equilibrium structure of alternating superconducting
and normal domains in a type-I superconductor is a complicated problem
that depends on, among other things, the nucleation history of 
the superfluid phase.  By flux conservation, the ratio of the 
sizes of the superfluid and normal domains is given by the 
relation $d_S/d_N = \sqrt{H_{\rm cm}/B}\sim 10$, where 
$B\sim 10^{12}$ G is the average value of the magnetic induction 
and $H_{\rm cm}\sim 10^{14}$ G is the thermodynamic critical magnetic field.

\begin{figure}[t] 
\begin{center}
\psfig{figure=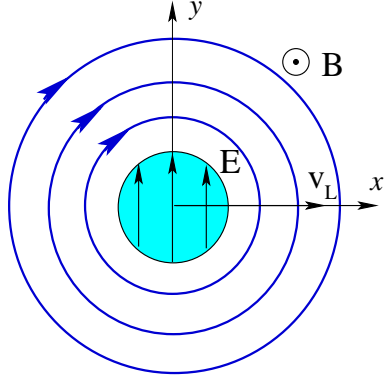,height=5.cm,width=5.cm,angle=0}
\end{center}
\caption{The structure of a rotational vortex placed 
in a type-I superconductor. The vortex velocity field is indicated
by the concentric circles. The non-superconducting domain (shaded region)
of radius $a$ is coaxial with the vortex and carries a magnetic 
field $H_{\rm cm}\sim 10^{14}$ G.  The vortex motion along the 
$x$-axis generates a transverse electric field, which drives 
the electron current through the domain and causes Ohmic 
dissipation.
}
\label{fig:DOMAIN}
\end{figure}

We first examine a model\cite{SSZ} in which the magnetic field
generated by the entrainment effect supports the formation of domains
coaxial with the neutron vortex.  Consider a vortex that moves 
at a constant velocity $\vecv_L$ and carries a coaxial normal 
domain of protonic fluid relative to the background electron liquid. 
Continuity of the electrochemical potentials of the superfluid 
and normal phases across the boundary between them entails the
existence of a constant transverse electric field 
\be\label{eq:6} 
E  = -\frac{m_p^*v_L v_p(a)}{ea}
\ee
across the normal domain (see Fig.~\ref{fig:DOMAIN}). 
The energy dissipated per unit length of a vortex is  
$W = \sigma E^2\, (a/b)^2$, where $\sigma$ is the electrical 
conductivity and the factor $(a/b)^2$ is the fractional 
area occupied by the domain.
Combining this relation with Eq.~(\ref{eq:6}), 
we obtain an alternative expression $W = \eta v_L^2$ 
for the dissipation, which identifies the friction 
coefficient as\cite{TYPEI}
\be\label{eq:7}  
\zeta = \frac{\eta}{\rho_S\omega_S} = \frac{\sigma}
{\rho_S\omega_Sc^2}
\left(\frac{\Phi_1}{2\pi ab}\right)^2
\left[\frac{a}{\lambda}{\rm ln}\left(\frac{b}{a}\right)
{\rm coth}\left(\frac{b-a}{a}\right)-1
\right]^2\,,
\ee
where $\Phi_1 = (\rho_{12}/\rho_{11})\Phi_0$.
The zero-field conductivity of ultra-relativistic
electrons is $\sigma_0 = n_e e^2 c\tau_c/p_{eF}$,
where $p_{eF}$ is the electron Fermi momentum, $n_e$ is the 
electron number density, and $\tau_c$ is 
the relaxation time for Coulomb scattering of electrons off 
protons in the normal domains.\cite{BAYM_NATURE}  The conductivity 
$\sigma = {\sigma_0}/{(\omega_c\tau_c)^2}$ entering Eq.~(\ref{eq:7}) 
includes the effect of the bending of electron trajectories in the 
magnetic field.  In this formula, $\omega_c = eH_{cm}/p_F$ 
is the electron cyclotron frequency, which is proportional to
the thermodynamic critical field $H_{cm}$.  For the typical density 
range $\rho = (7.9-8.6) \times 10^{14}$ g/cm$^3$, the friction coefficient 
has the order of magnitude $\zeta \simeq 10^{-5}$ at the temperature 
$T = 10^8$ K and scales with 
temperature as $\zeta \propto T^{2}$.  The drag-to-lift ratio 
satisfies the condition $\zeta \ll 1$ for all $T$ below
the critical temperature of the superfluid phase transition.

We turn next to the friction coefficient as predicted by the
model of Buckley et~al.,\cite{BUCKLEY} in which the normal domains 
contain a large number $N \sim 10$ neutron vortices.  In this case,
the damping of the differential rotation between the electron-proton 
plasma and the neutron superfluid is due to the interaction of 
domain (i.e., non-superconducting) protons with the core 
quasiparticles confined in the neutron-vortex core.  The relaxation 
process is therefore the same as in the case where the proton fluid is 
non-superconducting over the entire bulk of the core.  However, 
the result needs to be rescaled by the ratio of the areas occupied 
by the normal and superconducting layers.  The relaxation 
time per single vortex is\cite{SEDRAKIAN98}
\be\label{eq:10} 
\tau_{np} = 6
\left(\frac{p_{Fp}}{p_{Fn}}\right)^4\frac{m_n\mu_{pn}^*}
{\hbar m_p^*T\sigma_{np}}
\,{\rm exp}\left(0.02 \frac{\Delta_n^2}{\epsilon_{Fn}T}\right)\,,
\ee
where $p_{Fp}$ and $p_{Fn}$ are the Fermi momenta of protons and 
neutrons, $\mu_{pn}^* = m_p^* m_n^*/(m_p^* +m_n^*)$ is the reduced 
effective mass in terms of the proton and neutron effective masses,  
$\sigma_{np}$ is the total in-medium neutron-proton scattering
cross section, $\Delta_n$ is the gap in the neutron quasiparticle spectrum, 
and $\epsilon_{Fn}$ is the neutron Fermi energy. [Eq.~(\ref{eq:10})
differs from the analogous expression in Ref.~\refcite{SEDRAKIAN98} by
the factor $4m_n/\hbar P$; here $P$ is the pulsar period and
$m_n$ is the free-space neutron mass.]

In the relaxation-time approximation and zero-temperature 
limit, the friction is written as 
\be
\zeta  = \frac{\eta}{\rho_S\omega_S} 
= \frac{\hbar p_{Fp} n_p}{\rho_S\omega_Sc\tau_{np}}\,,
\ee
where $n_p$ is the proton number density.  For proton densities 
in the range $\rho_p = (4-8) \times 10^{14}$ g/cm$^3$ and temperatures 
$T \sim 10^{7-8}$, one finds $\zeta \le 0.1$.
For a given model of the type-I superconducting structure, 
the friction coefficient $\zeta$ must be rescaled by 
a factor $(d_N/d_S)^2\sim 0.01$. 

Now we are in a position to discuss the implications of the 
friction mechanisms described above for 
neutron-star precession.  The condition $(I_S/I_N)\zeta < 1$ 
seems to fulfilled unless $I_S/I_N \gg 1$.  The magnitude of 
the ratio $I_S/I_N$ depends on the superfluid/normal-fluid 
friction within all superfluid regions of the neutron star and 
is hard to assess.  Glitches and post-glitch relaxation 
provide a model-independent lower bound, $I_S/I_N \ge 0.1$. 
On the other hand, an upper bound is difficult to set.  The 
deep interior of the star, if superfluid, could be decoupled 
from the observable parts of the star on evolutionary time scales, 
without any effect on short-time-scale physics 
(although one does require $\zeta \to 0$, rather than 
$\zeta\to\infty$, to prevent damping of the precession). 
At any rate, it is rather unlikely that $I_S/I_N$ exceeds unity by 
many orders of magnitude, and appealing to the lower bound 
on the ratio of the moments of inertia, one can conclude 
that the precession is undamped for both dissipation 
mechanisms considered.

\section{Concluding remarks}
\label{SECTION6}

This review has covered a number of aspects of nucleonic 
superfluidity, ranging from microscopic theories of pairing 
in nuclear systems and neutron stars to mesoscopic frictional processes 
in superfluids and rotational anomalies in pulsars. Our survey of
this important subfield is by no means complete.  While the 
selection of topics is naturally biased toward the primary 
interests of the authors, we have chosen topics and problems 
with the intent of elucidating (i) the fascinating relationships 
between the physics involved at different scales and (ii) the 
richness of the contributions from diverse subfields of physics.
We close by listing a number of issues and problems that 
call for further clarification and concerted effort within 
the general framework of our discussions.
\begin{itemize}
\item 
The pairing problem at the level of mean-field BCS theory,
with the pairing driven directly by in-vacuum nuclear interactions,
is essentially solved within the density range over which these
interactions are constrained by experiment.  On the other hand,
issues such as the screening of nuclear interactions, 
renormalization of the single-particle spectrum, and off-shell 
energy behavior of the pairing gap still defy quantitative 
resolution.  Broadly speaking, extensions beyond the BCS theory 
are needed that incorporate fluctuation corrections while 
providing a consistent treatment of short-range correlations.

\item Superfluid phases with broken space-time symmetries have received
much attention from theorists in recent years, while experimental 
realization of asymmetric superfluids opens the possibility of 
testing the predictions of theory.  Importantly, relevant experiments 
are now probing the BCS-BEC crossover via the Feshbach resonance 
mechanism.  Given a broad effort, there is the prospect of mapping
out the superfluid phase diagrams of interesting fermionic systems, both 
experimentally and theoretically, in the space of coupling strength, 
spin/isospin asymmetries, temperature, etc.

\item The rotational anomalies observed in neutron stars continue
to provide useful constraints on the state of the superfluid matter 
in neutron-star interiors.  Further theoretical studies of vortex 
dynamics, combined with pulsar timing observations, may be expected
to shed new light on the internal structure of the superfluid
phases of neutron stars, especially on the question whether 
protons form a type-I or type-II superconductor.
\end{itemize}

\section*{Acknowledgments}
\addcontentsline{toc}{section}{Acknowledgments}
We acknowledge useful interactions with A.~Bulgac, J.~M.~Cordes, 
W.~H.~Dickhoff, J.~Dukelsky, V.~A.~Khodel, B.~Link, U.~Lombardo, 
H.~M\"uther, A.~Polls, P.~Schuck, H.-J.~Schulze, I.~Wasserman, 
D.~N.~Voskresensky 
and M.~V.~Zverev. AS acknowledges research support through a 
Grant from the SFB 382 of the Deutsche Forschungsgemeinschaft; 
JWC, through Grant No.~PHY-0140316 from the U.S.\ National 
Science Foundation.


\begin{thebibliography}{99}
\bibitem{GLITCHES} S. L. Shemar and A. G. Lyne, 
                {\it Mon. Not. RAS} {\bf 282},   677 (1996). 
\bibitem{TIMING_NOISE} 	J. M. Cordes, G. S. Downs, and J. Krause-Polstorff, 
               {\it Astrophys. J.} {\bf 330},  847 (1988).
\bibitem{PRECESSION1} 	I. H. Stairs, A. G. Lyne, and S. L. Shemar,
                         {\it Nature}  {\bf 406}, 484 (2000).
\bibitem{XRAYS1}  D. N. Voskresensky, {\it Lecture Notes in Physics}  vol.
{\bf 578} (Springer-Verlag, New York, 2001) pg. 467. 
\bibitem{XRAYS2}	D. G. Yakovlev and C. J. Pethick, 
	{\it Ann. Rev. Astron. Astrophys.} {\bf 42},  169 (2004).
\bibitem{XRAYS3} D. Page, U. Geppert, and F. Weber, {\it Nucl. Phys. A.},
                      in press.
\bibitem{XRAYS4} A. Sedrakian, {\it Prog. Part. Nucl. Phys.}, in press, 
                 {\tt nucl-th/0601086}.
\bibitem{GRAVITY1} L. Lindblom and G. Mendell,  {\it Phys. Rev.} {\bf D 61}, 4003 (2000).
\bibitem{GRAVITY2} A. Sedrakian and I. Wasserman, {\it Phys. Rev.}
                   {\bf D 63},  024016 (2000).
\bibitem{GRAVITY3} N. Andersson, G. L. Comer, and K. Grosart,         
            {\it Mon. Not. RAS} {\bf 355},  918 (2004). 
\bibitem{MIGDAL} A. B. Migdal, {\it Zh. Eksp. Teor. Fiz.} {\bf 37},  249  (1959)
                 [{\it Sov. Phys. JETP} {\bf 10},  176 (1960).]
\bibitem{PINES} A. Bohr, B. Mottelson, and D. Pines, {\it Phys. Rev.} {\bf 110},
                 936 (1958).
\bibitem{COOPER}	 L. N.	Cooper, R. L.  Mills, and A. M. Sessler, 
                    	{\it Phys. Rev.} {\bf 114}, 1377 (1959).  
\bibitem{BAYM_NATURE}	
	G. Baym, C. J. Pethick, and D. Pines, {\it Nature} {\bf  224}, 673 (1969).
\bibitem{YANG1}
C.-H.~Yang and J.~W.~Clark, {\it Nuovo Cimento Lett.} {\bf 3}, 272 (1970);
{\it ibid.}\ {\bf 2}, 379 (1970).
\bibitem{YANG2}
C.-H.~Yang and J.~W.~Clark, {\it Nucl. Phys.} {\bf A 174}, 49 (1971);
C.-H.~Yang Ph.D.~Thesis, Washington University in St.~Louis (1971).
\bibitem{CHAO} N.-C. Chao, J. W. Clark, and 
C.-H. Yang, {\it Nucl. Phys.} {\bf A179}, 320 (1972). 
\bibitem{HOFFBERG70}M. Hoffberg, A. E. Glassgold, R. W. Richardson, and 
                     M. Ruderman, {\it Phys. Rev. Lett.} {\bf 24},  775 (1970).
\bibitem{TAKATSUKA72} T. Takatsuka, 
{\it Prog. Theor. Phys.} {\bf 48},  1517 (1972).
\bibitem{P_WAVE0}  M. Baldo, O. Elgaroey,  
L. Engvik, M. Hjorth-Jensen, H.-J. Schulze,
     {\it Phys. Rev.}  {\bf C 58}, (1998) 1921.
\bibitem{P_WAVE} M. V. Zverev, J. W. Clark, V. A. Khodel
                      {\it Nucl. Phys.} {\bf A 720}, 20 (2003);
                      J. W. Clark, V. A. Khodel,
                      M. V. Zverev, nucl-th/0203046;
                      V. A. Khodel, J. W. Clark, M. V. Zverev,
                      {\it Phys. Rev. Lett.} {\bf 87}, 031103 (2001);
                      V. V. Khodel, V. A. Khodel, J. W. Clark, 
                      {\it Nucl. Phys.} {\bf A 679}, 827 (2001);
                      V. A. Khodel, J. W. Clark, M. Takano, and M. V. Zverev,
                      {\it Phys. Rev. Lett.} {\bf 93}, 151101 (2004).
\bibitem{D_WAVE1}  T. Takatsuka and R. Tamagaki, {\it Prog. Theor. Phys. Suppl.}
                   {\bf 112}, 27 (1993).
\bibitem{D_WAVE2} A. Sedrakian,  G. R\"opke, and
                   T. Alm, {\it Nucl. Phys.} {\bf 594}, 355 (1995);
                   T. Alm, G. R\"opke, A. Sedrakian, and F. Weber,
                   {\it Nucl. Phys.} {\bf A 604}, 491 (1996).
\bibitem{SDPAIRING1} T. Alm, G. R\"opke,
                       and M. Schmidt, {\it Z. Phys.} {\bf A 337}, 355 (1990).
\bibitem{SDPAIRING2}                    
  B. E. Vonderfecht, C. C. Gearhart, W. H. Dickhoff,
                       A. Polls, and  A. Ramos, {\it Phys. Lett.}
                       {\bf B 253},  1 (1991).
\bibitem{SDPAIRING3}   M. Baldo, I. Bombaci, and U. Lombardo, {\it Phys. Lett.}
                       {\bf B 283}, 8 (1992); 
M. Baldo, U. Lombardo, H.-J. Schulze, and Z. Wei
                       {\it Phys. Rev.} {\bf C 66}, 054304 (2002);
                       C. Shen, U. Lombardo, and P. Schuck
                       {\it Phys. Rev.}  {\bf C 71}, 054301 (2005).
\bibitem{SDPAIRING4}  H. M\"uther and W. H. Dickhoff, 
                      {\it Phys. Rev.}  {\bf C 72},  054313 (2005).
\bibitem{R1} U. Lombardo and H.-J. Schulze, 
                      {\it Lecture Notes in Physics, }
                      {\bf vol 578}, pg 30 (Springer, Berlin).
\bibitem{R2} D. J.  Dean and M. Hjorth-Jensen, {\it Rev. Mod. Phys.} {\bf 75},
             607 (2003).
\bibitem{ABRIKOSOV} A. A. Abrikosov, L. P. Gorkov, and I. E. Dzyaloshinski, 
{\it Methods of Quantum Field Theory in Statistical Physics}\ (Prienice-Hall,
Englewood Cliffs, NJ, 1963) (Dover, New York, 1975).
\bibitem{MIGDAL_TFFS}A. B. Migdal, {\it Theory of Finite Fermi-Systems}
(Nauka, Moscow, 1983, {\it in Russian}).
\bibitem{BALDO_Z} M. Baldo and A. Grasso, {\it Phys. Lett.} {\bf 485}, 115 (2000).
\bibitem{LOMBARDO_Z} U. Lombardo, P. Schuck, and W. Zuo, 
                      {\it Phys. Rev} {\bf C 64}, 021301 (2001).
\bibitem{KKC}
V. V. Khodel, V. A. Khodel, and J. W. Clark, {\it Nucl. Phys.} {\bf A598},
390 (1996).
%
\bibitem{TOM} A. Sedrakian, T. T. S. Kuo, H. M\"uther, and P. Schuck
              {\it Phys. Lett.}  {\bf  B 576}, 68 (2003).
\bibitem{PETHICK}  D. Pines and C. Pethick, in {\it Proc. XIth Intern. Conf.
                   on Low Temperature Physics}, ed. E. Kandu, (Kligatu
                   Publ. Co. Tokyo, 1971).
\bibitem{CLARK76} J. W. Clark, C. G. K\"allman, C. H. Yang, and 
                  D. A. Chakkalakal, {\it Phys. Lett.} {\bf B 61},  331 (1976).
\bibitem{LP1} S. Babu and G. Brown, {\it Ann. Phys.} {\bf 78}, 1 (1973).
\bibitem{LP2} S.-O. B\"ackman, G. E. Brown, and J. A. Niskanen, {\it Phys. Rep.}
               {\bf 124}, 1 (1985).
\bibitem{LP3} W. H. Dickhoff, A. Faessler, H. M\"uther, and S.-S. Wu, 
{\it Nucl.  Phys.} {\bf A 405},  534 (1983).
\bibitem{AINSWORTH} T. L. Ainsworth, J. Wambach, and D. Pines, 
                   {\it Phys. Lett.} {\bf B 222},  173 (1989).
\bibitem{WAMBACH} J. Wambach, T. L. Ainsworth,  and D. Pines, 
                   {\it Nucl. Phys.} {\bf A 555},  128 (1993).
\bibitem{CBF1} E. Feenberg, {\it Theory of quantum fluids} (Academic Press, NY 1969).
\bibitem{CBF2} J. W. Clark, {\it Prog. Part. Nucl. Phys.} {\bf 2},  89 (1979).
\bibitem{CBF3} J. W. Clark and P. Westhaus, {\it Phys. Rev.} {\bf 141},
 833 (1966).
\bibitem{CBF4} J. W. Clark, L. R. Mead, E. Krotscheck, K. E. K\"urten,
             and M. L. Ristig, {\it Nucl. Phys.} {\bf A 328}, 45 (1979).
\bibitem{CBF5} E. Krotscheck, R. A. Smith, and A. D. Jackson, {\it Phys. Rev.}
               {\bf B 24},  6404 (1981).
\bibitem{CHEN86} J. M. C. Chen, J. W. Clark, E. Krotscheck, and R. A. Smith,
                 {\it Nucl. Phys.} {\bf A 451}, (1986) 509. 
\bibitem{CHEN93} J. M. C. Chen, J. W. Clark, R. D. Dav\'e, and V. V. Khodel 
                  {\it Nucl. Phys.} {\bf A 555}, 59 (1993). 
\bibitem{FABROCINI} 	
	A. Fabrocini, S. Fantoni, A. Illarionov, and K. E. Schmidt, 
	{\it Phys. Rev. Lett.} {\bf 95}, 192501 (2005).
\bibitem{SCHULZE96} H.-J. Schulze, J. Cugnon, A. Lejune, M. Baldo, and 
                    U. Lombardo, {\it Phys. Lett.} {\bf B 375}, 1 (1996).
\bibitem{SCHWENK} A. Schwenk, B. Friman, and G. Brown, {\it Nucl. Phys.}
               {\bf A 713},  191 (2003).
\bibitem{WEISE}  T. Ericson and W. Weise,  {\it Pions and Nuclei} 
                  (Claredon Press, Oxford, 1988).
\bibitem{SEDRAKIAN03}A. Sedrakian, {\it Phys. Rev.}  
               {\bf C 68},  065805 (2003).
\bibitem{ELIASHBERG}G. M. Eliashberg, {\it Zh. Eksp. Teor. Fiz.} 
{\bf 38}, 966 (1960)
[{\it Sov. Phys. JETP} {\bf 11}, 696 (1960)].
\bibitem{KHODEL} V. A. Khodel,  J. W. Clark, M. Takano, and M. V. Zverev,
                 {\it Phys. Rev. Lett.} {\bf 93},  151101 (2004).
\bibitem{KRO_CLARKII}
E. Krotscheck and J. W. Clark, {\it Nucl.~Phys.} {\bf A 328}, 73
(1979).
\bibitem{KRO_CLARKIII}
E. Krotscheck and J. W. Clark, {\it Nucl.~Phys.} {\bf A 333}, 77
(1980).
\bibitem{WIRINGA1}
R.~B.~Wiringa, V.~G.~J.~Stoks, and R.~Schiavilla, {\it Phys.~Rev.}
{\bf C 51}, 38 (1995).
\bibitem{FANTONI81}
S. Fantoni, {\it Nucl.~Phys.}~{\bf A 363}, 381 (1981).
\bibitem{FANTONI_ROSATI}
S. Fantoni and S. Rosati, {\it Nuovo Cim.} {\bf 25A}, 593 (1975).
\bibitem{FRIMAN}
A. Schwenk and B. Friman, {\it Phys. Rev. Lett.} {\bf 92}, 082501 (2004).
\bibitem{TAKATSUKA81}
T. Takatsuka and R. Tamagaki, {\it Prog. Theor. Phys} {\bf 65}, 1333 (1981). 
\bibitem{CLOGSTON}A. M. Clogston, {\it Phys. Rev. Lett.} {\bf 9}, 266 (1962).
\bibitem{CHANDRA}B. S. Chandrasekhar, {\it Appl. Phys. Lett.} {\bf 1}, 7 (1962).
\bibitem{SARMA}G. Sarma, {\it  Phys. Chem. Solids} {\bf 24}, 1029 (1963).
\bibitem{GR}  L. P. Gor'kov and A. I. Rusinov,  {\it Zh. Eksp. Teor. Fiz. }
              {\bf 46}, 1363 (1964) [{\it Sov. Phys. JETP} {\bf 19}, 922 (1964)].
\bibitem{SAL}  A. Sedrakian, T. Alm, and U. Lombardo, 
                 {\it Phys. Rev.}  {\bf C 55}, R582 (1997).
\bibitem{SL00} A. Sedrakian and U. Lombardo, {\it Phys. Rev. Lett.} 
               {\bf 84}, 602 (2000).
\bibitem{ALFORD_GAPLESS} M. Alford, J. Berges, and K. Rajagopal, 
                       {\it Phys. Rev. Lett.} {\bf 84},  598 (2000).
\bibitem{RUSTER} I. A. Shovkovy, S. B. Ruester, 
                 D. H. Rischke, {\it J. Phys. G} {\bf 31},  849 (2005)
                    and refs. therein.
\bibitem{JAIKUMAR} P. Jaikumar, C. D. Roberts, and A. Sedrakian, 
                 {\it Phys. Rev.} {\bf C 73}, 042801 (2006).
\bibitem{LO}A. I. Larkin and Yu. N. Ovcihnnikov,  
            {\it Zh. Eksp. Teor. Fiz.} {\bf 47},  1136 (1964) 
            [{\it Sov. Phys. JETP} {\bf 20}, 762 (1965)].
\bibitem{FF} P. Fulde and R. A. Ferrell, 
             {\it Phys. Rev.} {\bf 135}, A550 (1964).
\bibitem{BOWERS_RAJAGOPAL} J.~A.~Bowers and K.~Rajagopal,
                          {\it Phys.\ Rev.}  {\bf D 66}, 065002 (2002).
\bibitem{SEDRAKIAN00} A. Sedrakian, {\it Phys. Rev.}  {\bf C 63},  025801 (2001).
\bibitem{Casalbuoni:2003jn}R.~Casalbuoni, R.~Gatto,
                           M.~Mannarelli, and G.~Nardulli,
                           {\it Phys.\ Rev.}  {\bf D 66}, 014006 (2002).
\bibitem{DFS1}      H. M\"uther and A. Sedrakian, {\it Phys.\ Rev.\ Lett.} 
                    {\bf 88}, 252503 (2002);  
{\it Phys.\ Rev.}  {\bf C 67},   015802 (2003).
\bibitem{DFS2} A. Sedrakian, J. Mur-Petit, A. Polls, and H, M\"uther, 
              {\it Phys. Rev.}  {\bf A 72},  013613 (2005).
\bibitem{CALDAS} P.~F. Bedaque, H.~ Caldas, and G.~ Rupak, 
             {\it Phys. Rev. Lett.} {\bf 91}, 247002 (2003); 
	     H.~Caldas, {\it Phys. Rev. } {\bf A 69}, 063602 (2004)
\bibitem{COMB}  R.~Combescot, {\it Europhys. Lett} {\bf  55},  15 (2001).
\bibitem{GUB}  E. Gubankova, E.G. Mishchenko, and F. Wilczek, 
                 {\it Phys. Rev. Lett.} {\bf 94}, 110402 (2005).
\bibitem{LEG} A. J. Leggett, 
 in {\it Modern Trends in the Theory of Condensed Matter} 
 (Springer, Berlin, 1980), p.13;
 {\it J. Phys. (Paris)} {\bf 41}, (1980) C7-19.
\bibitem{KEL} L. V. Keldysh and Yu. V. Kopaev, 
{\it Sov. Phys. Solid State} {\bf 6},  2219 (1965); 
 L. V. Keldysh and A. N. Kozlov, 
{\it  Sov. Phys. JETP} {\bf 27},   521 (1968).
\bibitem{NOZ} P. Nozi\`eres and S. Schmitt-Rink, 
 {\it J. Low Temp. Phys.} {\bf 59},  195 (1985).
\bibitem{ALM}  
 T. Alm, B. L. Friman, G. R\"opke, and H. Schulz,  
 {\it Nucl. Phys.} {\bf A 551}, 45 (1993).
\bibitem{SCK}  M. Baldo, U. Lombardo, and P. Schuck,
 {\it Phys. Rev.} {\bf C 52}, 975 (1995).
\bibitem{SSALR}  H. Stein, A. Schnell, T. Alm, and G. R\"opke,
{\it  Z. Phys.} {\bf A 351}, 295 (1995). 
\bibitem{LNSSS}U. Lombardo, P Nozi\`eres, P. Schuck, H.-J. Schulze,
       and  A. Sedrakian, {\it Phys. Rev. } {\bf C 64}  064314 (2001).
\bibitem{SC} A. Sedrakian and J. W. Clark, 
             {\it Phys. Rev.} {\bf C 73},  035803 (2006). 
\bibitem{FESHBACH1} 
W.~C.~Stwalley, {\it Phys. Rev. Lett.} {\bf 37}, 1628 (1976).
\bibitem{FESHBACH2} E.~Tiesinga, B.~J.~Verhaar, and H.~T.~C.~Stoof,
  {\it Phys. Rev. } {\bf   A47},   4114 (1993).
\bibitem{FESHBACH3} Ph. Courteille, R.~S. Freeland, D.~J. Heinzen, 
F.~A. van Abeelen, and B.~J. Verhaar,  {\it Phys. Rev. Lett.} {\bf  81},  69 (1998).
\bibitem{ZWI} W. Zwierlein, A. Schirotzek, C. H. Schunck, W. Ketterle,
                  {\it Science} {\bf 311},  492 (2006).
\bibitem{HULET} G. B. Partridge, W. Li, R. I. Kamar, Y. 
                Liao, R. G. Hulet, {\it Science} {\bf 311}, 503 (2006).
\bibitem{ATOMS_THEORY1}
J.~Mur-Petit, A.~Polls, and H.-J.~Schulze, {\it Phys. Lett.}   
                       {\bf  A 290},   317 (2001);
C. Mora and R.~Combescot, {\it Physica} {\bf B 329},   1435 (2003);
W. V. Liu and F. Wilczek,   {\it Phys. Rev. Lett.} {\bf 90},  047002 (2003);
 M. M. Forbes, E. Gubankova, W. V. Liu, and F. Wilczek
{\it Phys. Rev. Lett.} {\bf 94},  017001 (2005);
W. Yi and L.-M. Duan, {\it Phys. Rev.}  {\bf A 73}, 031604(R) (2006);
A. Bulgac, M. M. Forbes, and A. Schwenk, eprint cond-mat/0602274;
Kun Yang and S. Sachdev, {\it Phys. Rev. Lett.} {\bf 96}, 187001 (2006);
              S. Sachdev and Kun Yang, Phys. Rev. {\bf B 73}, (2006) 174504; 
A. Sedrakian, H. M\"uther, and A. Polls, eprint cond-mat/0605085.
\bibitem{ATOMS_THEORY2}
Kun Yang, {\it {\it Phys. Rev.} Lett.} {\bf 95}, 218903 (2005); 
	       e-print cond-mat/0508484; e-print cond-mat/0603190;
A. Sedrakian, J. Mur-Petit, A. Polls,
                and  H. M\"uther, {\it Phys. Rev.}  {\bf A 72},  013613 (2005);
J. Dukelsky, G. Ortiz, and S.~M.~A. Rombouts, {\it Phys. Rev. Lett.} 
                                {\bf 96}, 180404 (2006).
\bibitem{ATOMS_THEORY3}
T. Mizushima, K. Machida, and M. Ichioka, {\it Phys. Rev. Lett.} 
                           {\bf 94},  060404 (2005);
P. Castorina, M. Grasso, M. Oertel, M. Urban, and
                   D. Zappala,  {\it Phys. Rev.}  {\bf A 72},  025601 (2005);
F. Chevy, eprint cond-mat/0601122; 
T. N. De Silva and E. J. Mueller, {\it Phys. Rev.} {\bf A 73}, 
     051602(R) (2006);  eprint cond-mat/0604638.
\bibitem{VS}  G. A. Vardanyan and D. M. Sedrakyan, {\it Sov. Phys. JETP}
                   {\bf 54}, 919 (1981).
\bibitem{MENDELL_LINDBLOM}   G. Mendell and L. Lindblom,
                 {\it Ann. Phys.} {\bf 205}, 110 (1990)
\bibitem{MENDELL} G. Mendell, {\it Astrophys. J.} {\bf 380},  515 (1991); 
                      ibid, pg 530.
\bibitem{SS95} A. Sedrakian and D. Sedrakian,  
               {\it Astrophys. J.} {\bf 380},  515 (1991).
\bibitem{MENDELL2}  G. Mendell,
	Mon. Not. RAS {\bf 296},  903 (1998).
\bibitem{PRIX} R. Prix,	{\it Phys. Rev.}  {\bf D 62},  3005 (2000)
\bibitem{ANDERSSON} N. Andersson, T. Sidery, and G. L.  Comer, 	
        {\it Mon. Not. RAS} {\bf 368},  162 (2006).
\bibitem{ALPAR1} M. A. Alpar, S. J. Langer, and J. A. Sauls, 
                {\it Astrophys. J.} {\bf 282}, 533 (1984).
\bibitem{RUDERMAN}	M. Ruderman, 
	{\it Astrophys. J.} {\bf 382},  5 (1991).
\bibitem{SS1} D. M. Sedrakian and K. M. Shahabasian, {\it Sov. Phys. Uspekhi}
              {\bf 34} 555 (1991) and references therein.
\bibitem{MUZIKAR} P. Muzikar and C. J. Pethick, 
              {\it Phys. Rev.}  {\bf  B 24}, 2533 (1981).
\bibitem{SED_CORDES} A. Sedrakian and J. Cordes, 
                     {\it Mon. Not. RAS }
                     {\bf 307}, 365 (1999).
\bibitem{SSZ} D. M. Sedrakian, A. Sedrakian, and G. F. Zharkov, 
              {\it Mon. Not. RAS} {\bf 290}, 203 (1997);
             {\it Comptes Rendus Acad. Sci. Paris Ser. IIb }
              {\bf 325}, 763 (1997). 
\bibitem{GZ} V. L. Ginzburg and G. F. Zharkov, {\it Journ. Low. Temp. Phys.}
                  {\bf 92}, 25 (1993).
\bibitem{BUCKLEY} K. B. W. Buckley, M. A. Metlitski, and A. R. Zhitnitsky, 
                  {\it  Phys. Rev. Lett.} {\bf 92}, 151102 (2004); 
{\it Phys. Rev.}   {\bf  C 69}, 055803 (2004). 
\bibitem{ALFORD} M. Alford, G. Good, and S. Reddy, 
                   {\it Phys. Rev.}  {\bf C 72},  055801 (2005).
\bibitem{ALPAR2}	 M. A.	Alpar,  H. F. Chau,  K. S. Cheng, 
                 and D. Pines, 
                    {\it  Astrophys. J.} {\bf 409},  345 (1993).
\bibitem{LINK}	 B. Link, R. I. Epstein, and G. Baym, 
           	{\it Astrophys. J.} {\bf 403},  285 (1993).

\bibitem{SHAHAM} J. Shaham, Astrophys. J. {\bf 214}, (1977) 251.
\bibitem{SWC} A. Sedrakian, I. Wasserman, and J. M. Cordes,
                  {\it Astrophys. J.} {\bf 524}, 341 (1999).
\bibitem{BLINK} B. Link, {\it Phys. Rev. Lett.} {\bf 91}, 101101 (2003).
\bibitem{WASSERMAN} I. Wasserman, {\it Mon. Not. RAS }
                 {\bf 341}, 1020 (2003).
\bibitem{AKGUN} T. Akgun, B. Link, and I. Wasserman,
                        {\it  Mon. Not. RAS} {\bf 365},  653 (2006).
\bibitem{FEIBELMAN} P. J. Feibelman, {\it Phys. Rev.}  {\bf D 4}, 1589 (1971).
\bibitem{SAULS} P. Muzikar, J. A. Sauls, and J. W.  Serene,
                {\it Phys. Rev.} {\bf D 21}, 1494 (1980).
\bibitem{SEDRAKIAN98} A. Sedrakian, {\it Phys. Rev.}  {\bf D 58}, 021301(R) (1998).
\bibitem{CHENG} K. Y. Ding, K. S.  Cheng, and H. F. Chau, 
                        {\it Astrophys. J.}   {\bf 408}, 167 (1993).
\bibitem{REPINNING} A. Sedrakian, 
                {\it  Mon. Not. RAS} {\bf 277},  225 (1995).
\bibitem{BAYM_EPSTEIN}G. Baym and R. Epstein, {\it Astrophys. J.} {\bf 387}, 276 (1992).
\bibitem{JONES}	 P. B.	Jones, {\it Mon. Not. RAS} {\bf 257}, 501 (1992).
\bibitem{TYPEI} A. Sedrakian, {\it Phys. Rev.}  {\bf D 71},  083003 (2005).
\end{thebibliography}
\end{document}